\documentclass[pra,twocolumn,preprintnumbers,amsmath,amssymb,showpacs]{revtex4-1}
\usepackage{graphicx}
\usepackage{amssymb}
\usepackage{mathrsfs}
\usepackage{color}
\usepackage{bbold}

\newcommand{\beq}{\begin{equation}}
\newcommand{\eeq}{\end{equation}}

\begin{document}

\title{ Topological order, symmetry, and Hall response of two-dimensional spin-singlet superconductors}

\author{Sergej Moroz$^{1, 2,3}$, Abhinav Prem$^{1}$, Victor Gurarie$^{1}$, and Leo Radzihovsky$^{1,3}$}
\affiliation{$^1$Department of Physics and Center for Theory of Quantum Matter,\\
                 University of Colorado, Boulder, Colorado 80309, USA \\
             $^2$Department of Physics, Technical University of Munich, 85748 Garching, Germany\\
             $^3$Kavli Institute for Theoretical Physics, University of California, Santa Barbara, California 93106, USA}


\begin{abstract}
Fully gapped two-dimensional superconductors coupled to dynamical electromagnetism are known to exhibit topological order. In this work, we develop a unified low-energy description for spin-singlet paired states by deriving topological Chern-Simons field theories for $s$-wave, $d+id$, and chiral higher even-wave superconductors. These theories capture the quantum statistics and fusion rules of Bogoliubov quasiparticles and vortices and incorporate global continuous symmetries - specifically, spin rotation and conservation of magnetic flux - present in all singlet superconductors. For all such systems, we compute the Hall response for these symmetries and investigate the physics at the edge. In particular, the weakly-coupled phase of a chiral $d+id$ chiral state has a spin Hall coefficient $\nu_s=2$ and a vanishing Hall response for the magnetic flux symmetry. We argue that the latter is a generic result for two-dimensional superconductors with gapped photons, thereby demonstrating the absence of a spontaneous magnetic field in the ground state of chiral superconductors. It is also shown that the Chern-Simons theories of chiral spin-singlet superconductors derived here fall into Kitaev's 16-fold classification of topological superconductors.
\end{abstract}


\maketitle


\section{Introduction} \label{intro}

Topological order is a cornerstone of contemporary quantum condensed matter physics \cite{wenbook, Fradkin2013}. Rooted in the discovery of the fractional quantum Hall effect, it arises in numerous phases of matter such as spin liquids and fractional quantum Hall liquids. Topological order manifests itself in the fractionalization of low-energy excitations, ground state degeneracy on closed manifolds, and long-range entanglement \cite{Zeng2015}.

The idea that \emph{two-dimensional superconductors}, i.e., electrically charged paired fermions that couple to a \emph{dynamical} two-dimensional electromagnetic field, are topologically ordered has been appreciated for some time \cite{Kivelson1990, Wen1991a, Balents1998, Read2000, Hansson2004}. 
Due to the Higgs mechanism, there are no Goldstone modes in the energy spectrum and, at zero temperature, the superconductor is \emph{fully gapped}. There are two types of \emph{point} excitations in a two-dimensional superconductor: \emph{Bogoliubov quasiparticles} and \emph{vortices}.  At energies much lower than the gap, only the topological properties of these low-energy excitations (their braiding and fusion rules) and their symmetry quantum numbers matter. This information can be effectively encoded in a \emph{topological} field theory. For an $s$-wave superconductor, a topological Chern-Simons theory  was elucidated and described in detail in a beautiful paper by Hansson, Oganesyan, and Sondhi \cite{Hansson2004}.

More recently, attempts have been made to construct topological theories for non-abelian $p+ip$ superconductors \cite{Hansson2012, Hansson2015}. These states are an important example in the field of two-dimensional fermionic \emph{chiral} superfluidity and superconductivity, which has been the focus of experimental and theoretical condensed matter physics for decades. Today, chiral $p+ip$ pairing plays a central role in research fields as diverse as the physics of $^3$He \cite{vollhardt, volovikbook}, quantum Hall physics \cite{Read2000}, unconventional superconductivity \cite{Kallin2015}, cold atoms \cite{Gurarie2005, Gurarie20072}, and topological quantum computation \cite{nayak2008}. 

Our focus in this paper is on \emph{chiral spin-singlet} paired states which have also received some attention in the past. For example, in the weakly-coupled \emph{abelian} topological phase, which falls into class C of the ten-fold way classification \cite{ryu2010}, a chiral $d+id$ superconductor is predicted to exhibit a spin Hall effect \cite{Senthil1999, Read2000}, support four protected chiral edge modes \cite{Teo2010}, and have a non-universal edge mass current carried by unpaired fermions \cite{Tada2015, Volovik2014a}. Over the past few years, new physical motivations necessitating further study of chiral $d+id$ pairing  have emerged. Specifically, two-dimensional materials with a \emph{hexagonal} lattice symmetry necessarily have degenerate $d_{x^2-y^2}$ and $d_{xy}$ gaps \cite{Kallin2015}, which makes them good candidates for chiral $d+id$ superconductors. Two well-known examples where $d+id$ pairing is currently believed to be relevant are the pnictide $\text{SrPtAs}$ \cite{Nishikubo2011, Fischer2014} and graphene doped to the van Hove filling \cite{nandkishore2012, Honerkamp2014, Kiesel2012}. Chiral spin-singlet superconductors paired in higher partial even-waves are also theoretically interesting and are likely to become experimentally relevant in the future.

In a two-dimensional chiral superconductor, the spontaneous breaking of parity and time-reversal symmetries might lead one to expect a \emph{spontaneous} generation of a finite magnetic field in the ground state, which originates from the internal motion of fermions orbiting around each other in Cooper pairs. For a type II superconductor, this would result in a finite density of quantum vortices in the ground state.  One might thus anticipate a relation between the density of elementary fermions, $n_f$, and the density of vortices, $n_v$, of the form $n_v=\nu_v n_f$. An important question that motivated this work is whether $\nu_v$ is a universal quantized number for chiral paired states such as a $d+ id$ superconductor. 
 
To answer this question and to extend the understanding of superconductors as topologically ordered states, we present a general framework for studying the low-energy physics of spin-singlet superconductors.  We start from the \emph{microscopic} theory of these states that are spin-rotationally invariant and conserve magnetic flux. From there, we \emph{derive} topological field theories for different \emph{gapped} abelian states, i.e., $s$-wave, $d+id$, and chiral superconductors paired in higher partial even-waves. The virtue of this approach is that it naturally encodes the braiding and fusion rules of low-energy excitations, incorporates symmetries, and captures the physics at the edge.  In the case of chiral states, we reproduce the known spin Hall effect \cite{Senthil1999, Read2000}. Moreover, we investigate the vortex Hall response associated with the magnetic flux symmetry. The coefficient $\nu_v$, introduced above, equals the Hall coefficient for this response. Importantly, we demonstrate explicitly that for all superconductors considered here, $\nu_v$ is \emph{zero}. As a result, we predict that there is no \emph{ spontaneous magnetic field}, and thus no dense array of vortices in the ground state of a chiral two-dimensional superconductor. By merging the ideas discussed above, our work unifies topological order and symmetries in superconductors and firmly establishes them as \emph{symmetry enriched topological} (SET) phases \cite{Chen2013}.

It is worth remembering that in this work the electromagnetic field is assumed to be strictly confined to \emph{two} spatial dimensions. Is such flatland electromagnetism actually realizable in an experiment? It is clear that without special arrangements electromagnetism will penetrate into the third dimension, and as a result, one has to deal with a \emph{mixed dimension} problem where paired fermions are confined to two spatial dimensions but electromagnetism is three-dimensional. By embedding two dimensional electrons inside a specific dielectric medium, however, electromagnetism can also be made \emph{effectively} two-dimensional. 
To this end, one can engineer a two-dimensional logarithmic potential between electric charges
by surrounding the sample by a low-permittivity medium with a dielectric constant $\epsilon_{\text{medium}}\ll \epsilon_{\text{sample}}$ \cite{Keldysh1979}. Secondly, the magnetic field lines can be arranged to
be transverse to the boundary of the quasi-two dimensional sample by sandwiching
it between a high-permeability ($\mu_{\text{medium}}\gg \mu_{\text{sample}}$) material. Thus, at least static electromagnetism can be effectively confined to two dimensions.

The outline of our paper is as follows: In Sec. \ref{WZ}, we first introduce our framework -- an abelian Chern-Simons field theory that captures the topological properties, symmetries, and edge physics of abelian gapped states. Next, in Sec. \ref{modelref}, we introduce topologically ordered spin-singlet superconductors and identify their internal global symmetries. Then, in Sec. \ref{microsec}, we  derive the topological theories of $s$-wave and $d+id$ superconductors by starting from the low-energy model of a non-relativistic $d_{x^2-y^2}$ paired state. In that section, we also extend this derivation to chiral superconductors paired in higher partial even-waves. In Sec. \ref{TFTs}, we analyze the resulting Chern-Simons theories. Here, we first devote Sec. \ref{swave} to a conventional ($s$-wave) superconductor. In Sec. \ref{dwave}, we then investigate the effective theory of a $d+id$ superconductor and calculate its spin and vortex Hall responses. Next, in Sec. \ref{higher}, we describe the extension of our construction to chiral spin-singlet superconductors paired in higher partial even-waves and demonstrate that these fall into the sixteen-fold way classification of chiral superconducting states developed by Kitaev \cite{Kitaev2006}. Sec. \ref{noHall} presents a general argument that elucidates why chiral superconductors have a zero vortex Hall coefficient. Finally, in Sec. \ref{disc}, we close with some open questions that go beyond the scope of this paper.

\section{Abelian topological field theories} \label{WZ}
 The low-energy physics of a completely gapped two-dimensional state of matter is encoded in a topological field theory.
Moreover, since the spin-singlet superconductors studied in this paper are known to form only \emph{abelian phases} \footnote{An \emph{abelian phase} is characterized by a unique anyon resulting from the fusion of any two excitations. On the other hand, \emph{non-abelian} anyons can fuse into different outcomes.}, we propose that topological aspects of  such phases can be captured by an abelian Chern-Simons field theory \cite{Read1990, Wen1992b, Frohlich1991}
\beq \label{EFT}
\begin{split}
 \mathcal{L}_{\text{bulk}}=&\frac{1}{4\pi} \epsilon^{\mu\nu\rho} a^I_{\mu} K_{IJ}\partial_\nu a^J_{\rho}
 - a^I_{\mu} j_{I}^{\mu} 
 -\frac{1}{2\pi} t_{AI} \epsilon^{\mu\nu\rho}\mathcal{A}^A_{\mu}\partial_\nu a^I_\rho.
 \end{split}
\eeq 
Here $a^I$ is a multiplet ($I=1,2,\dots, N$) of auxiliary statistical gauge fields \footnote{A remark regarding \emph{compactness}: The statistical fields dual to global conserved currents are non-compact which ensures absence of instanton magnetic monopoles and strict conservation of these currents. On the other hand, compact statistical gauge fields are also present sometimes, but these do not encode any strict conservation laws.}, $K_{IJ}$ is a symmetric \emph{integer-valued} $N\times N$ matrix that determines the self and mutual statistics of excitations, and $j_I$ are quasiparticle currents. Note that $a_\mu$ are coupled to \emph{quantized} charges carried by the currents $j_I$. As a result, the first-quantized current densities are $j_I^0(\mathbf{r})=\sum_n l_I^{(n)} \delta(\mathbf{r}-\mathbf{r}^{(n)})$, characterized by an integer-valued gauge charge vector $l^{(n)}$ and the position $\mathbf{r}^{(n)}$ of the $n^{\text{th}}$ quasiparticle excitation. 
The third term in Eq. \eqref{EFT} represents the coupling to external sources $\mathcal{A}^A$ of $A=1,2,\dots, M$ \emph{global} U$(1)_A$ symmetries. The theory \eqref{EFT} has proven to be successful in describing the low-energy properties of abelian quantum Hall fluids \cite{Wen1995}. 

Importantly, \emph{topological order} is simply encoded in the effective theory \eqref{EFT}. Indeed, the ground state degeneracy on a torus, a direct manifestation of topological order, is fixed by the determinant of the $K$-matrix \cite{Wen1995}
\beq
\# \text{GS}=|\text{det} K|.
\eeq 
Moreover, this determinant also fixes the number of independent anyon types (see, for example, \cite{Lu2012}).
 
While the effective field theory \eqref{EFT} is quite an inefficient formalism for encoding the fusion and braiding rules of the bulk excitations \footnote{Indeed, the number of entries of the $K$-matrix might be much larger than the number of independent braiding phases.}, it is in fact very well suited for understanding the physics of the edge. Following Wen \cite{Wen1992a}, in the absence of external sources $\mathcal{A}^A$, one finds a chiral Luttinger theory of $N$ chiral bosons $\phi^I$ propagating along the edge 
\beq \label{edge}
\mathcal{L}_{\text{edge}}=\frac{1}{4\pi}\Big[K_{IJ}\partial_t \phi^I \partial_x \phi^J-V_{IJ}\partial_x \phi^I \partial_x \phi^J \Big].
\eeq
Here $V_{IJ}$ is a non-universal positive-definite real matrix that depends on the microscopic properties of the edge. 

Systems with a finite chiral central charge $c$ at the edge have a non-zero thermal Hall conductance and host $c$ \emph{co-propagating} bosonic edge modes that cannot be gapped by backscattering.
In particular, for chiral spin-singlet superconductors, with the chirality parameter $k$ to be defined in Eq. \eqref{gap}, one finds $c=k$.

A natural next question to ask is whether \emph{counter-propagating} edge modes can be gapped without breaking any symmetries or if they are symmetry-protected , i.e., are stable against arbitrary symmetry preserving local perturbations. In the rest of this section, we present a brief analysis of this question.

To understand the structure of a straight edge along the $x$ direction in the presence of external sources $\mathcal{A}^A$, we start from Eq. \eqref{EFT} in the absence of quasiparticle currents $j_I^\mu$. Following \cite{Wen1992a}, we first impose a gauge fixing condition in the bulk 
\beq \label{Gauge}
a_t^I + v^I a_x^I = 0.
\eeq
The Gauss law constraint (or incompressibility condition) 
\beq
2\pi\frac{\delta S}{\delta a_t^I} = K_{IJ} b^J - t_{AI} \mathcal{B}^A = 0,
\eeq
where $b^I = \epsilon^{ij}\partial_i a^I_j$ and $\mathcal{B}^A = \epsilon^{ij}\partial_i \mathcal{A}^A_j$, is automatically satisfied by
\beq \label{Gauss}
a_i^I = \partial_i \phi^I + K^{-1}_{IJ} t_{AJ}\mathcal{A}_i^A.
\eeq
Substituting Eqs. \eqref{Gauge} and \eqref{Gauss} into the bulk action \eqref{EFT}, with some manipulations, will result in the generalization of the effective edge action \eqref{edge} in the presence of sources $\mathcal{A}^A$. For the purposes of understanding the fate of counter-propagating edge modes in the presence of symmetries, however, it suffices to work out the transformation properties of the edge fields $\phi_I$. Under a local U$(1)_A$ transformation parametrized by $\alpha_A$, the background gauge fields and the edge multiplet transform as
\beq \label{symtrans}
\begin{split}
\delta \mathcal{A}^A_\mu&=-\partial_\mu \alpha_A, \\
\delta \phi_I&=\alpha_A K^{-1}_{IJ} t_{AJ}. 
\end{split}
\eeq
The last equation follows from Eq. \eqref{Gauss} and the U$(1)_A$ neutrality of all statistical gauge fields $a^I$.
We consider now an edge perturbation of the form
\beq
\int dx\,dt\, 
\cos(l_I \phi_I).
\eeq
First, the requirement of locality of this term enforces that $\vec{l}\in \mathbb{Z}^N$ must be bosonic (have trivial self and mutual statistics) \cite{Lu2012}. Moreover, it follows from Eq. \eqref{symtrans} that these terms (sometimes called ``Higgs'' terms) are invariant under all global symmetries if
$
l^{T}\cdot K^{-1}\cdot t_A = 0 
$ 
for all $A=1,\dots, M$.
In addition, according to the null vector condition of \cite{Haldane1995}, such symmetry allowed Higgs terms can now gap a pair of counter-propagating edge modes if and only if
$
l^{T}\cdot K^{-1}\cdot l = 0
$ 
since the two fields can then be rotated such that they form a single non-chiral Luttinger liquid that is gapped by backscattering \cite{Levin2013}. More generally, for gapping $n$ pairs of counter-propagating edge modes, we will require $n$-independent (commuting) Higgs terms that can simultaneously provide energy gaps to all of these edge modes. 

In summary, in the presence of U$(1)$ global symmetries, $n$ pairs of counter-propagating edge modes can be gapped if and only if one can find $\vec{l}_i \in \mathbb{Z}^N (i = 1,\cdots,n)$ such that
\begin{subequations} \label{gapconds}
\begin{itemize}
\item The Higgs terms are constructed from elementary bosonic excitations:
\beq
2 \pi l_i^T\cdot K^{-1}\cdot l'=0\quad(\text{mod } 2 \pi)\quad \forall i, \forall \vec{l'}\in\mathbb{Z^N}
\eeq
\item The Higgs terms are charge neutral under all global symmetries:
\beq
l_i^T\cdot K^{-1}\cdot t_A=0 \quad \forall i, \forall A
\eeq
\item The null vector conditions are satisfied:
\beq
l_i^T\cdot K^{-1}\cdot l_j=0\quad \forall i,j
\eeq
\end{itemize}
\end{subequations}
As shown later, all systems considered in this paper host at least one pair of counter-propagating modes which are gapped out by the Higgs terms satisfying the above conditions.

 \section{Spin-singlet superconductors: topological order and symmetries}  \label{modelref}
In this paper, we consider \emph{two-dimensional} electrically charged spinful fermions which couple to a \emph{dynamical} electromagnetic gauge field that is also confined strictly to two spatial dimensions. We will \emph{assume} that, due to electromagnetism and some spin-independent short-range attractive interaction, the fermions pair in a \emph{spin-singlet} chiral channel with the gap
 \beq \label{gap}
 \Delta_\mathbf{p}=(p_x\pm i p_y)^k \Delta_0
 \eeq
where the sign defines the chirality and $k$ is an even integer due to antisymmetry of the fermionic singlet pair wave-function. In fact, $k$ is just the orbital angular momentum carried by a Cooper pair. 

We will discuss separately a conventional $s$-wave superconductor ($k=0$), $d+id$ superconductor ($k=2$), and higher partial even-wave chiral superconductors ($k=4,6,\dots$). 
 The explicit construction and analysis of the topological Chern-Simons theories of these superconductors depends on the chirality parameter, $k$, and will appear in separate sections below. Here, we first highlight the generic properties that all of these systems have in common:
 \begin{itemize}
 \item {\bf Topological order:} Two-dimensional spin-singlet superconductors with a \emph{dynamical} gauge field exhibit \emph{topological order}. For an $s$-wave superconductor this has been emphasized in \cite{Hansson2004}, where the ground state degeneracy on a torus was found to be equal to four. In fact, this result also holds for the $d+id$ superconductor \cite{Read2000} and can be easily extended to higher partial wave spin-singlet chiral pairing. As a result, all superconductors considered in this paper have $|\text{det} K|=4$ and contain four independent anyons, which we call $1$, $e$, $m$, and $\epsilon$, following a common convention. As we will see in the following, topological order in superconductors leads to fractionalization of the quantum numbers and statistics of low-energy excitations. 
 \item {\bf Internal continuous global symmetries:} In a superconductor, electromagnetism, being a \emph{gauge redundancy}, is \emph{not} a global symmetry. 
There are, however, two internal global symmetries of spin-singlet superconductors to be considered in this paper:  First, the spin-singlet structure of the pairing implies that a non-abelian SU$(2)_s$ spin rotation is a global symmetry. Since by construction, the effective theory \eqref{EFT} can couple only to \emph{abelian} sources, here we consider the Cartan subalgebra of SU$(2)_s$ with the charge $Q_s\sim S_z$ and introduce in Eq. \eqref{EFT} an external abelian spin source $\mathcal{A}^s$ that couples to the z-component of the spin current. We will see that in superconductors, this charge is carried only by Bogoliubov quasiparticles while vortices are spinless. Second, \emph{any} two-dimensional superconductor has a global abelian U$(1)_v$ \emph{magnetic flux} symmetry with a charge $Q_v=\int d^2x B$ \cite{Kovner1991}. The conservation of this charge follows from the electromagnetic Bianchi identity (Faraday's law) $\epsilon^{\mu\nu\rho}\partial_{\mu}F_{\nu\rho}=0$, which is valid provided there are no magnetic monopoles. This appears naturally in a model where the electromagnetic U$(1)$ gauge group is \emph{non-compact}, which we consider in this paper.  
In a type-II superconductor, a vortex carries one-half of a magnetic flux quantum, i.e., a $\pi$-flux, while a Bogoliubov quasiparticle is \emph{neutral} with respect to this symmetry. As a result, in a superconductor the flux charge defined above is carried only by vortices. Correspondingly, in the effective theory \eqref{EFT} we introduce an external abelian source, $\mathcal{A}^v$, which couples to the charge $Q_v$. It is worth emphasizing that due to the presence of a gap, neither of the two global symmetries introduced above are broken spontaneously in a superconducting ground state.
 \end{itemize}

 \section{Derivation of Chern-Simons topological field theories of spin-singlet superconductors}  \label{microsec}
 
We start from the low energy model of a weakly-coupled non-relativistic $d_{x^2-y^2}$ superconductor and, by deforming it, will derive the topological Chern-Simons theories for $s$-wave and $d+id$ superconductors. At the end of this section, we will also extend this derivation to spin-singlet chiral superconductors that are paired in higher partial even-waves. 

Before presenting the derivation, it is worth noting that for a relativistic $s$-wave superconductor a topological Chern-Simons theory was derived in \cite{Hansson2004}. In contrast to that construction, our derivation applies to all gapped spin-singlet non-relativistic superconductors and includes coupling to external magnetic flux and spin symmetry sources.

Our starting point is the Lagrangian of a gapless two-dimensional $d_{x^2-y^2}$ superconductor \footnote{\label{fa} For simplicity we set the mass of an elementary fermion to unity. Its electric charge $e$ is set to \emph{minus} unity, which fixes the magnetic flux carried by an elementary (counterclockwise) vortex $\varphi(\mathbf{x})=\text{arg}(\mathbf{x})/2$ to $+\pi$, i.e., a half of a magnetic flux quantum.}
\beq \label{micro}
\begin{split}
\mathcal{L}^0=&-\frac{1}{4} F_{\mu\nu}F^{\mu\nu}-n_s \mathcal{D}_t \varphi +\frac {n_s} {2} \frac{1}{c_s^2} (\mathcal{D}_t \varphi)^2-\frac{n_s}{2} (\mathcal{D}_i \varphi)^2 \\
&-A_{\mu}j^{\mu}_{\text{ions}}+\mathcal{L}^0_{\text{qp}}(\psi_i, A; \mathcal{A}^s)-\frac {1}{2\pi}\epsilon^{\mu\nu\rho} \mathcal{\tilde A}^{v}_{\mu}\partial_{\nu} A_\rho,
\end{split}
\eeq
where the covariant derivative $\mathcal{D}_\mu \varphi=\partial_\mu \varphi-A_\mu$, $n_s$ is the superfluid density, 
and $c_s$ is the speed of sound. The first term \footnote{Here the indices are raised and lowered with the Minkowski metric.} in Eq. \eqref{micro} encodes the Maxwell dynamics of the electromagnetic field $A$ and the next three terms incorporate the dynamics of the fluctuating part, $\varphi$, of the superconductor phase. 
In addition, we have included the neutralizing ion static background that carries the electromagnetic current $j^{\mu}_{\text{ions}}= n_f \delta^{\mu 0}$, where $n_f$ is the density of elementary fermions that undergo pairing. $\mathcal{L}^0_{\text{qp}}$, specified later in this section, incorporates the low-energy physics of gapless spinful fermionic quasiparticles that couple to electromagnetism, $A$, and to the spin source, $\mathcal{A}^s$. Finally, the last term in Eq. \eqref{micro} describes the coupling of the magnetic flux symmetry current to its external source, $\mathcal{\tilde A}^v$. Its prefactor fixes the magnetic flux charge of an elementary counterclockwise vortex to $Q_v=+1/2$ (see footnote \cite{Note4}). 

\subsection{Vortices}
In the presence of vortices the superconductor phase $\varphi$ can be split into the regular part $\varphi_{\text{reg}}$ and the vortex part $\varphi_{\text{v}}$, which has singularities at the positions of vortices. By a suitable regular gauge transformation, the regular part $\varphi_{\text{reg}}$ can be absorbed into the electromagnetic potential, $A_\mu$ (Higgs mechanism). On the other hand, the singular part $\varphi_{\text{v}}$ determines the \emph{conserved} vortex current
\beq \label{cond1}
j_{\text{v}}^{\mu}=\frac{1}{\pi}\epsilon^{\mu\nu\rho}\partial_\nu \partial_\rho \varphi_{\text{v}}.
\eeq 
After introducing the statistical gauge field $a_{\mu}\equiv \partial_\mu \varphi_{\text{v}}$, that is dual to the vortex current, the Lagrangian \eqref{micro} becomes
\beq \label{micro1}
\begin{split}
\mathcal{L}^0=&-\frac{1}{4} F_{\mu\nu}F^{\mu\nu}- b_{\mu}\Big(j_{\text{v}}^{\mu}-\frac 1 \pi \epsilon^{\mu\nu\rho} \partial_\nu a_\rho \Big)\\
&-n_s (a_t-A_t) +\frac {n_s} {2} \frac{1}{c_s^2} (a_t-A_t)^2-\frac{n_s}{2} (a_i-A_i)^2 \\
&-A_{\mu}j^{\mu}_{\text{ions}}+\mathcal{L}^0_{\text{qp}}(\psi_i, A; \mathcal{A}^s)-\frac {1}{2\pi}\epsilon^{\mu\nu\rho} \mathcal{\tilde A}^{v}_{\mu}\partial_{\nu} A_\rho,
\end{split}
\eeq
where following \cite{Hansson2004} we introduced the (Lagrange multiplier) statistical gauge field $b_{\mu}$, whose equation of motion enforces the condition \eqref{cond1}. 

\subsection{Bogoliubov quasiparticles}
\begin{figure}[ht]
\begin{center}
\includegraphics[width=0.3\textwidth]{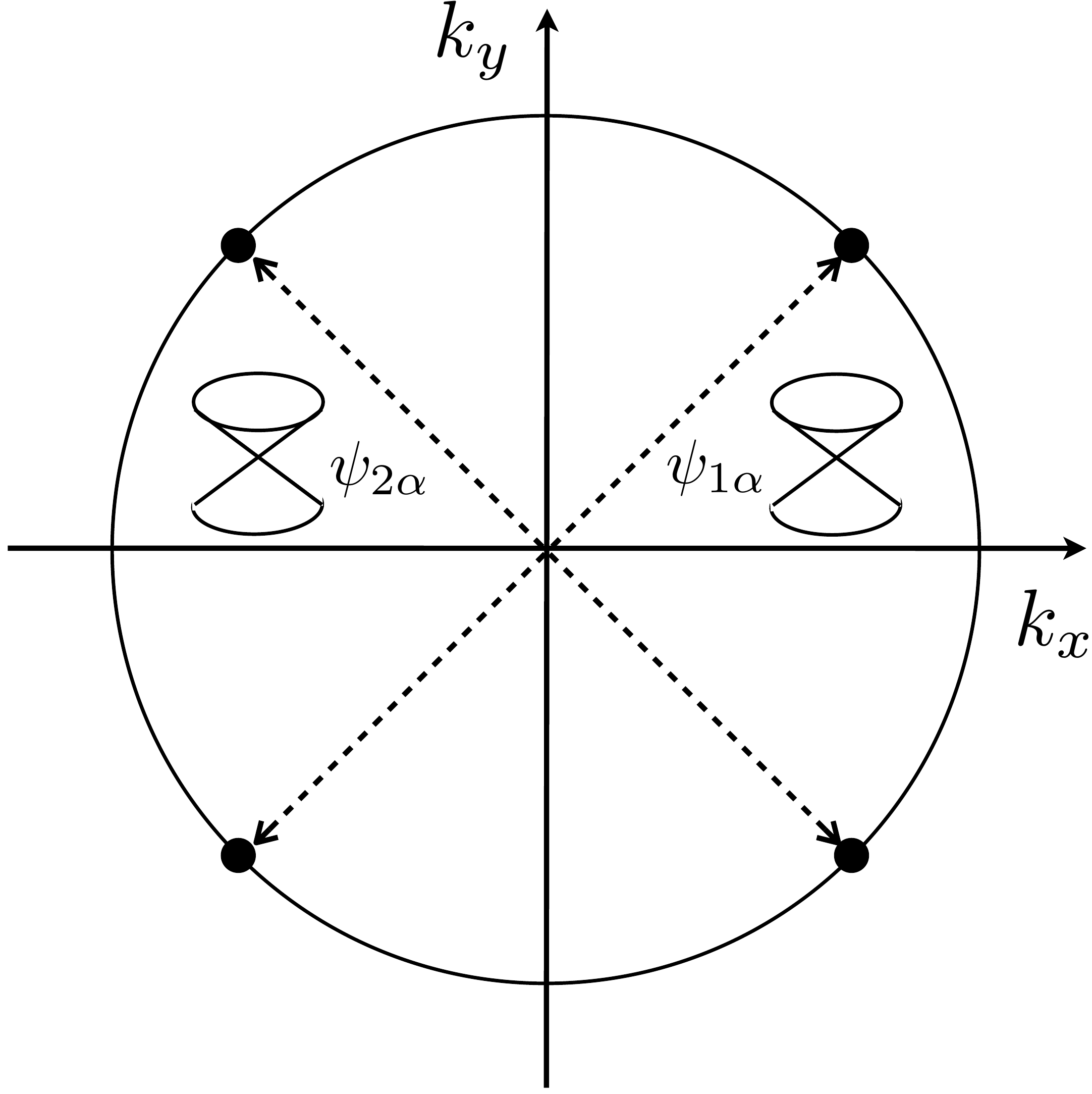}
\caption{Four Dirac fermions $\psi_{i\alpha}$ arising from the two pairs of nodes (connected by dashed lines) at the Fermi surface of a $d_{x^2-y^2}$ superconductor.}\label{fig2}
\end{center}
\end{figure}

We now specify the quasiparticle Lagrangian, $\mathcal{L}^0_{\text{qp}}$. The $d_{x^2-y^2}$ superconductor has four nodes on the Fermi surface (see Fig. \ref{fig2}), where Bogoliubov quasiparticles become gapless. After linearizing near these nodes one finds \emph{four} massless two-component Dirac modes $\psi_{i\alpha}$, where we introduced the nodal index $i=1,2$ and the spin index $\alpha=\uparrow, \downarrow$. The low-energy Dirac Lagrangian of these quasiparticles was derived in \cite{Balents1998, Balents1999} and also discussed in \cite{Hermanns2008}. In the absence of electromagnetism it is given by
\beq \label{dxy}
\begin{split}
\mathcal{L}^0_{\text{qp}}=&\psi_1^\dagger i \partial_t \psi_1+
\psi_1^\dagger \Big(i v_F \partial_X \tau_z+ i v_\Delta \sum_{s=+,-} e^{is\varphi} \partial_Y e^{is\varphi} \tau_s \Big)\psi_1 \\
&+(1\leftrightarrow 2, X\leftrightarrow Y),
\end{split}
\eeq
where $\tau_{\pm}=(\tau_x\pm i \tau_y)/2$.
Here we have introduced the node-aligned spatial coordinates $X=(x+y)/\sqrt{2}$ and $Y=(-x+y)/\sqrt{2}$ and suppressed both the spin index $\alpha$ and Dirac index in particle-hole space, where the Pauli matrices $\tau_i$ operate. The gap terms proportional to $\tau_+$ and $\tau_-$ contain the coupling to the superconductor phase $\varphi$. Note that our normalization of this phase differs from the normalization used in \cite{Balents1998, Balents1999} by a factor of two.

Since $\psi_{i\alpha}$ transforms as a doublet under the spin SU$(2)_s$ symmetry, the quasiparticles couple to the spin source $\mathcal{A}^s$ via the minimal coupling
$
\partial_\mu \to \partial_\mu +i \sigma_z \mathcal{A}^s_\mu/2 
$, where the Pauli matrix $\sigma_z$ acts in spin space. On the other hand, the Bogoliubov quasiparticle spinor $\psi_{i\alpha}$ is a combination of a particle and hole, and thus it is more subtle to introduce the electromagnetic potential $A$ into Eq. \eqref{dxy}. In fact, $A$ appears only in the kinetic terms (but not in the gap term) in Eq. \eqref{dxy} via the minimal coupling
$
\partial_\mu\to \partial_\mu +  i e \tau_z A_\mu
$, where we set $e=-1$ (see footnote \cite{Note4}).
In the absence of vortices it is customary at this point to transform to electrically neutral fermions, $\tilde \psi_i= \exp(-i \varphi \tau_z) \psi_i$, which eliminates the phase $\varphi$ from the gap term \cite{Balents1998}. 
Importantly, in the presence of vortices this transformation is not single-valued \footnote{Indeed, across the vortex branch cut, the phase of every Dirac component of the spinor $\tilde\psi_i= \exp(-i \varphi \tau_z) \psi_i$ changes by $\pm \pi$.}, and instead one should perform a single-valued transformation \cite{Anderson1998} (see also \cite{Ariad2015} for a general discussion)
\beq \label{tran}
\tilde \psi_i= \exp(-i \varphi \tau_z\pm i \varphi_{\text{v})} \psi_i.
\eeq
The presence of these neutral fermions, that braid trivially with all other quasiparticles, will be reflected in the fermionic nature of the $K$-matrices discussed in Sec.\ref{TFTs}.

After performing the transformation with the minus sign in Eq. \eqref{tran}, we find
\beq \label{dxy1}
\begin{split}
\mathcal{L}^0_{\text{qp}}(\tilde \psi_i, A; \mathcal{A}^s)=&\tilde \psi_1^\dagger \big[ i \mathcal{D}_t+\tau_z (A_t-a_t) \big] \tilde \psi_1\\
&+\tilde \psi_1^\dagger v_F \big[i \mathcal{D}_X \tau_z +(A_X-a_X)\big]\tilde \psi_1\\
&+\tilde \psi_1^\dagger v_\Delta  i\mathcal{D}_Y  \tau_x \tilde \psi_1 \\
&+(1\leftrightarrow 2, X\leftrightarrow Y),
\end{split}
\eeq
where
\beq \label{covderf}
\mathcal{D}_\mu=\partial_\mu+i a_\mu-i\sigma_z\mathcal{A}^s_\mu/2.
\eeq

\subsection{$s$-wave and $d+id$ deformations of $d_{x^2-y^2}$ superconductor}
\begin{figure}[ht]
\begin{center}
\includegraphics[width=0.3\textwidth]{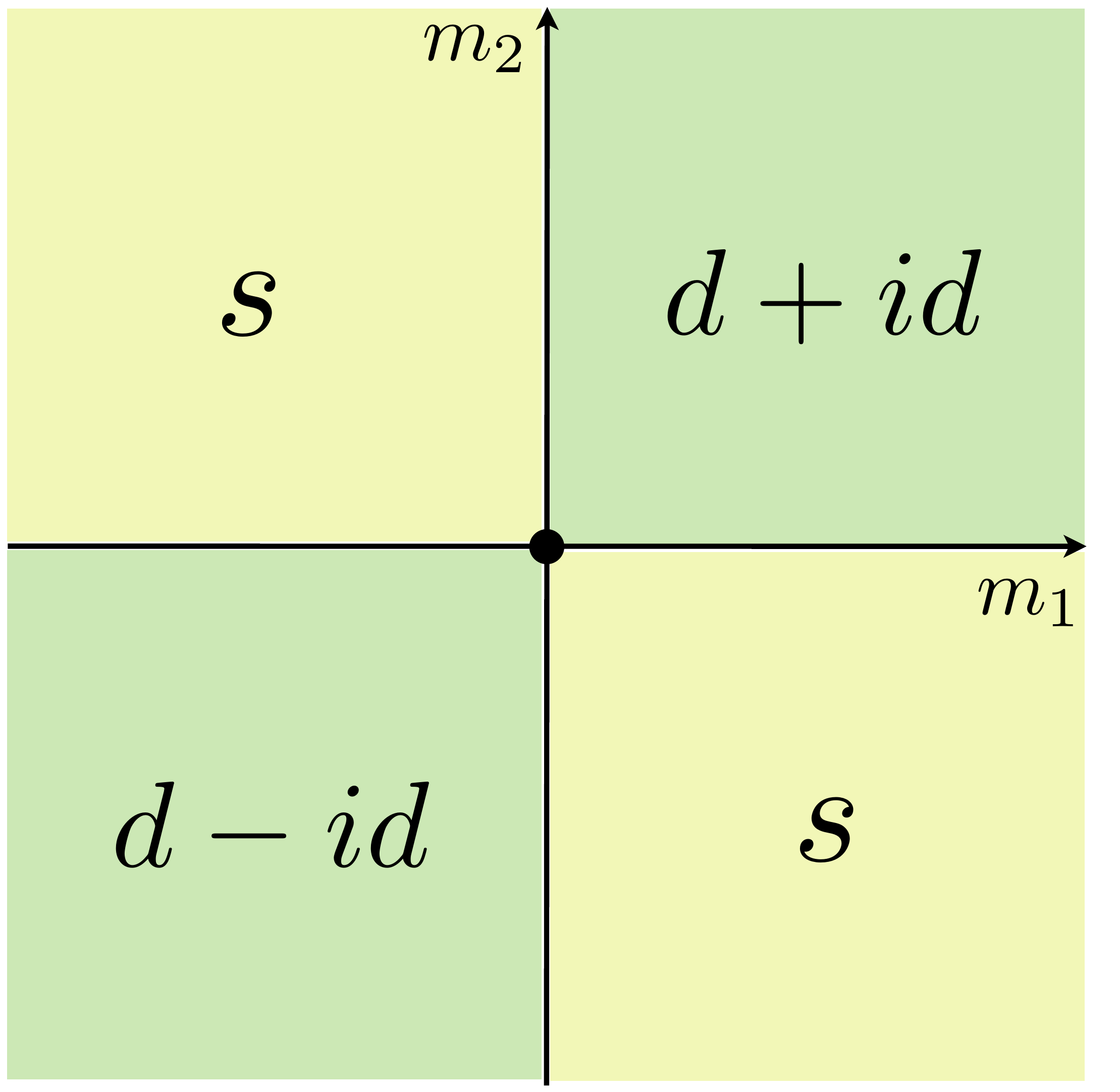}
\caption{Topological phase diagram of a deformed $d_{x^2-y^2}$ superconductor as a function of masses $m_i$ of the Dirac modes $\psi_i$. The origin represents the $d_{x^2-y^2}$ superconductor.}\label{fig3}
\end{center}
\end{figure}

Crucially for us, one can enter into the $s$-wave or $d+id$ gapped phase by adding appropriate masses to nodal Dirac quasiparticles that were introduced above. In particular, the terms \cite{Senthil1999}
\beq \label{deltaL}
\begin{split}
\mathcal{L}^{\delta}_{\text{qp}}&=-\delta (\tilde\psi_1^\dagger \tau_y \tilde\psi_1 \pm \tilde \psi_2^\dagger \tau_y \tilde \psi_2)\\
\end{split}
\eeq
add to the $d_{x^2-y^2}$ superconductor some amount of $is$ and $i d_{xy}$ pairing, respectively. Topologically, the resulting phases are equivalent to the $s$-wave and $d+id$ superconductors. We see thus that that Eq. \eqref{deltaL} is nothing but the mass term for the nodal Dirac particles \footnote{To this end, perform the following transformation of the last term in Eq. \eqref{dxy1}: first rotate by a $\pi/2$ angle, i.e., $X\to -Y$, $Y\to +X$ and second apply a unitary rotation $U= \exp(i \pi \tau_z)$ to $\tilde \psi_2$.}, with masses $m_1=\delta$ and $m_2=-\delta$. More generally, allowing arbitrary masses $m_i$ for the spinors $\psi_i$, one can show that their signs are the same ($m_1 m_2>0$) for the $d+id$ phase, while they are opposite ($m_1 m_2<0$) for the $s$-wave phase. The resulting phase diagram is summarized in Fig. \ref{fig3}. Since the $s$-wave and the $d+id$ superconductors differ only in the sign of the mass of the Dirac mode $\tilde \psi_2$, in the following we consider them in parallel.

\subsection{Integrating out electromagnetism}
Henceforth, we will assume that the Bogoliubov quasiparticles $\psi_i$ have a sufficiently large gap and thus carry an electric density and current that are negligible compared to the superfluid density and the supercurrent. As a result, the superfluid density $n_s$ is equal to the density $n_f$ of the elementary fermions. Given this, we first impose the charge neutrality condition $n_s+j^0_{\text{ions}}=0$ in Eq. \eqref{micro1}. Combining now Eqs. \eqref{micro1}, \eqref{dxy1} and \eqref{deltaL}, we integrate out the massive electromagnetic field $A_\mu$, which to lowest order in derivatives is equivalent to the substitutions $A_\mu \to a_\mu$. Keeping now only the leading terms in derivatives, we find
\beq \label{Lnew}
\begin{split}
\mathcal{L}\to &
 - b_{\mu}\Big(j_{\text{v}}^{\mu}-\frac 1 \pi \epsilon^{\mu\nu\rho} \partial_\nu a_\rho \Big)\\
&-n_s a_t +\mathcal{L}^{\mathcal{D}}_{\text{qp}}(\tilde \psi_i)-\frac {1}{2\pi}\epsilon^{\mu\nu\rho} \mathcal{\tilde A}^{v}_{\mu}\partial_{\nu} a_\rho
\end{split}
\eeq
with
\beq \label{Lqpf}
\begin{split}
\mathcal{L}^{\mathcal{D}}_{\text{qp}}(\tilde \psi_i)=&
 \tilde \psi_1^\dagger \Big(i \mathcal{D}_t+ iv_F \tau_z  \mathcal{D}_X+ i  v_\Delta \tau_x  \mathcal{D}_Y -\delta \tau_y \Big) \tilde\psi_1 \\
 + &\tilde \psi_2^\dagger \Big(i \mathcal{D}_t+ iv_F \tau_z  \mathcal{D}_X+ i  v_\Delta \tau_x  \mathcal{D}_Y \mp \delta \tau_y \Big) \tilde\psi_2,
 \end{split}
\eeq
where the covariant derivative $\mathcal{D}_\mu$ was defined in Eq. \eqref{covderf}.

Since in a superconductor an elementary fermion carries a $2\pi$ flux of the vortex magnetic field (see Appendix \ref{App1}), a finite density of these fermions gives rise to a finite background value of $\mathcal{B}^v$. It thus seems natural at this point to absorb the term $-n_s a_t$ in Eq. \eqref{Lnew} into the source term. This indeed can be done by writing
$
n_s a_t\to  \epsilon^{\mu\nu\rho}  \mathcal{\bar A}^{v}_\mu \partial_{\nu}a_{\rho}/(2\pi)
$
with $ \mathcal{\bar A}^{v}_t=0$ and $ \mathcal{\bar B}^v=2\pi n_s$. As the result, Eq. \eqref{Lnew} simplifies to
\beq \label{top}
\begin{split}
\mathcal{L}= & \frac 1 \pi \epsilon^{\mu\nu\rho} a_\mu \partial_\nu b_\rho- b_{\mu}j_{\text{v}}^{\mu}-\frac {1}{2\pi}\epsilon^{\mu\nu\rho} \mathcal{A}^{v}_{\mu}\partial_{\nu} a_\rho \\
&+\mathcal{L}^{\mathcal{D}}_{\text{qp}}(\tilde \psi_i),
\end{split}
\eeq
where $\mathcal{A}^{v}=\mathcal{\bar A}^{v}+ \mathcal{\tilde A}^{v}$.

\subsection{Topological field theory}
Finally, we integrate out massive Dirac fermions in Eq. \eqref{Lqpf}, which are minimally coupled to the statistical gauge field $a$ and the spin source $\mathcal{A}^s$. The resulting statistical and spin Hall response is
\beq
\mathcal{L}^{\mathcal{D}}_{res}=\frac{1}{8\pi} \sum_{\substack{i=1,2 \\ \alpha=\uparrow, \downarrow}} \frac{m_i}{|m_i|} \epsilon^{\mu\nu\rho} \Big(a_\mu\partial_\nu a_\rho+ q_s^2\mathcal{A}^s_\mu\partial_\nu \mathcal{A}^s_\rho\Big),
\eeq
where the unit of spin charge $q_s=1/2$.
This can be encoded within the Chern-Simons theories of two statistical gauge fields $c^\uparrow$ and $c^\downarrow$.

Specifically, for the $s$-wave case, where $\sum_{i,\alpha} m_i/|m_i|=0$, we use the zero-chirality theory
\beq \label{topfers}
\begin{split}
\mathcal{L}^{\mathcal{D}}_{\text{qp}}\to & \frac{1}{4\pi} \epsilon^{\mu\nu\rho} \big(c_\mu^\uparrow\partial_\nu c_\rho^\uparrow- c_\mu^\downarrow\partial_\nu c_\rho^\downarrow \big)\\
&+\frac{1}{2\pi}\epsilon^{\mu\nu\rho} a_\mu \partial_\nu (c^\uparrow_\rho+ c^\downarrow_\rho) \\
&-\frac{1}{2\pi}\epsilon^{\mu\nu\rho} \mathcal{A}^s_\mu \partial_\nu (c^\uparrow_\rho- c^\downarrow_\rho)\\
&-c^\uparrow_\mu j_{\uparrow}^\mu+ c^\downarrow_\mu j_{\downarrow}^\mu-\frac 1 \pi \epsilon^{\mu\nu\rho}\mathcal{A}^s_\mu \partial_\nu a_\rho
\end{split}
\eeq
which by construction has a vanishing Hall response. In particular, integrating out $c^{\uparrow}$ and $c^{\downarrow}$ results in a term $\sim \mathcal{A}^s \partial a$ which is exactly cancelled by the last term in Eq. \eqref{topfers}. 

On the other hand, for the $d+id$ case with $\sum_{i,\alpha} m_i/|m_i|=4$ we employ the chiral theory
\beq \label{topferdid}
\begin{split}
\mathcal{L}^{\mathcal{D}}_{\text{qp}}\to & \frac{1}{4\pi} \epsilon^{\mu\nu\rho} \big(c_\mu^\uparrow\partial_\nu c_\rho^\uparrow+ c_\mu^\downarrow\partial_\nu c_\rho^\downarrow \big)\\
&+\frac{1}{2\pi}\epsilon^{\mu\nu\rho} a_\mu \partial_\nu (c^\uparrow_\rho+ c^\downarrow_\rho) \\
&-\frac{1}{2\pi}\epsilon^{\mu\nu\rho} \mathcal{A}^s_\mu \partial_\nu (c^\uparrow_\rho- c^\downarrow_\rho)\\
&-c^\uparrow_\mu j_{\uparrow}^\mu- c^\downarrow_\mu j_{\downarrow}^\mu.
\end{split}
\eeq
In Eqs. \eqref{topfers} and \eqref{topferdid}, we also included the coupling to gapped spin-up and spin-down Bogoliubov quasiparticles that are represented in the Chern-Simons field theory by external bosonic currents $j_\uparrow$ and $j_\downarrow$, respectively \footnote{It would be instructive to derive Eqs. \eqref{topfers}, \eqref{topferdid} rigorously by using the functional bosonization approach developed in \cite{Chan2013}.}. By putting Eqs. \eqref{topfers}, \eqref{topferdid} that capture the quasiparticle sector of the full Lagrangian into Eq. \eqref{top} results in the completely topological \emph{four-component} $(a,b,c^{\uparrow}, c^{\downarrow})$ Chern-Simons theories for $s$-wave and $d+id$ superconductors which will be discussed in detail in Secs. \ref{swave} and \ref{dwave}, respectively.

\subsection{Extension to higher partial waves}
The above construction can be straightforwardly generalised to chiral superconductors paired in the $k^{\text{th}}$ partial wave, for $k\in2\mathbb{Z}/\{0\}$. We start from a time-reversal invariant superconductor with a gap proportional to the real part of $(p_x\pm ip_y)^k$ which has $k$ pairs of nodes at the Fermi surface. This leads to $k$ massless Dirac spin doublets. We can deform into the chiral $(p_x+ip_y)^k$ state by adding masses of the same sign to all Dirac modes. This procedure gives rise to the Chern-Simons theory that will be investigated in Sec. \ref{higher}, following our presentation and analysis for the $s$-wave and $d+id$ cases.
  
 \section{Topological field theories of spin-singlet superconductors} \label{TFTs}
 In this section we analyze the topological theories that were derived in the previous section.
 In addition to summarizing the braiding properties of quasiparticles, we investigate the role of global symmetries in the bulk and at the edge of spin-singlet superconductors.
 

\subsection{$s$-wave superconductor} \label{swave}

Having encoded the topological properties of vortices in Eq. \eqref{top} and of the nodal Dirac quasiparticles in Eq. \eqref{topfers}, we combine these to arrive at the four-component Chern-Simons theory of the form \eqref{EFT} with $a^I=(a, b, c^{\uparrow}, c^{\downarrow} )$. The resulting $K$-matrix characterizing the $s$-wave state is
\beq \label{Ks}
K=\left(
\begin{tabular}{ cccc}
  0 & 2 &1 & 1 \\
  2 & 0 &0 & 0  \\
  1 & 0 &1 & 0 \\
  1 & 0 &0 & -1 
\end{tabular}
\right).
\eeq

\begin{figure}[ht]
\begin{center}
\includegraphics[height=0.25\textwidth]{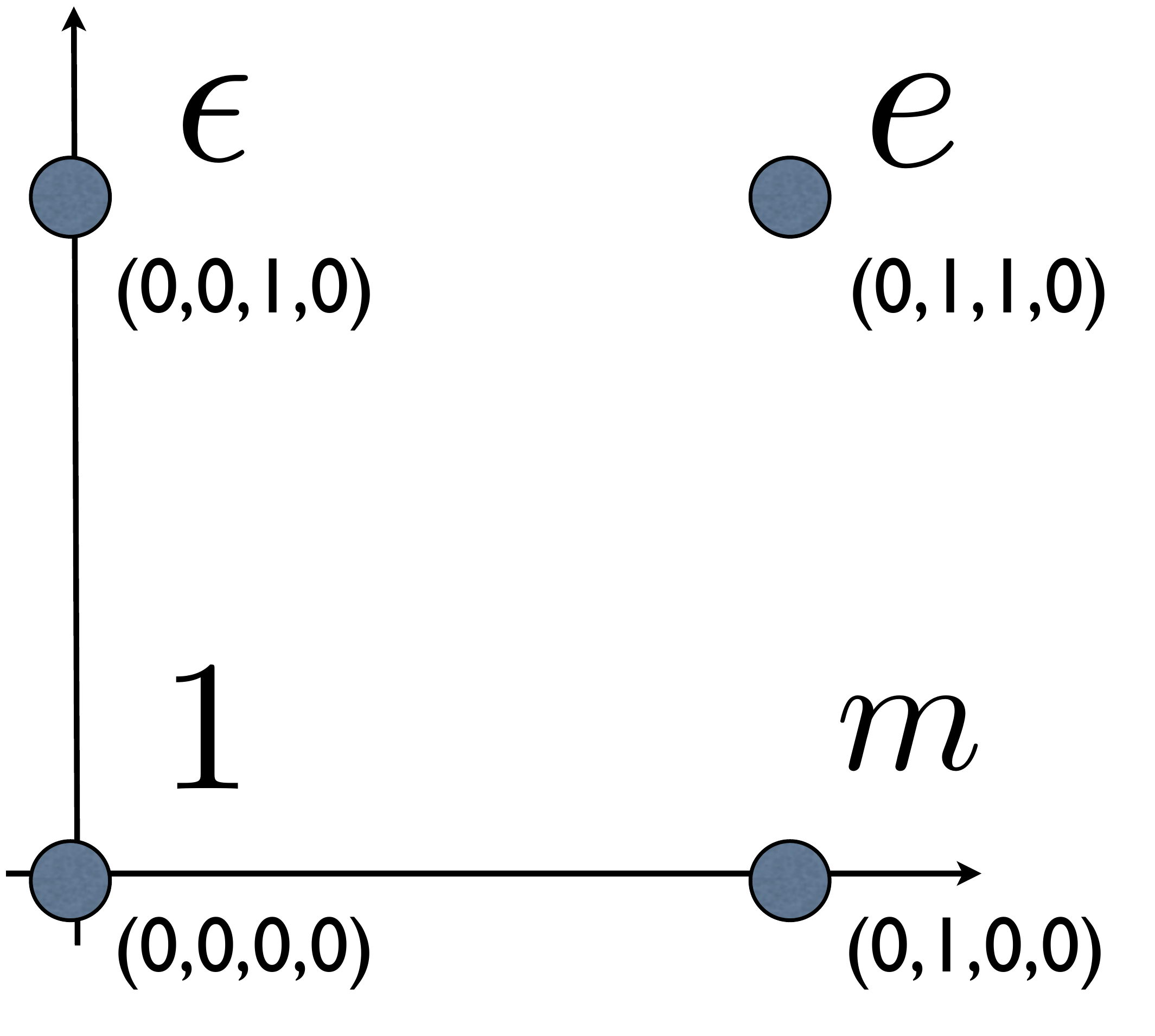}
\caption{Integer-valued $l$-vectors of $1$, $e$, $m$, $\epsilon$ excitations for the $s$-wave and $d+ id$ superconductor. }\label{fig1a}
\end{center}
\end{figure}

This $K$-matrix is \emph{fermionic} (with two odd integers on the diagonal) encoding the presence of an \emph{elementary} \footnote{By definition, an elementary excitation has trivial mutual statistics with all anyons.} fermionic excitation in the spectrum of the superconductor. Notably, the $K$-matrix \eqref{Ks} does contain the (toric code) bosonic block $\big(
\begin{tabular}{ cc}
  0 & 2 \\
  2 & 0  \\
\end{tabular}
\big)$, which, for the $s$-wave superconductor, was derived previously in \cite{Hansson2004}.
The $l$-vectors, defining integer-valued charges of independent excitations with respect to statistical gauge fields $a, b, c^\uparrow, c^{\downarrow}$, are shown \footnote{Specifically, in Fig. \ref{fig1a} we chose to identify $\epsilon$ with the spin-up Bogoliubov quasiparticle that carries the current $j_{\uparrow}$ in Eq. \eqref{topfers}. Alternatively, in the $s$-wave case one can choose $l_{\epsilon\downarrow}=(0,0,0, -1)$ which corresponds to the spin-down Bogoliubov quasiparticle. The latter identification gives the same braiding and fusion rules as the former, but obviously differs by the sign of the spin charge. Note that for the $d+id$ state, Eq. \eqref{topferdid} leads to $l_{\epsilon\downarrow}=(0,0,0, 1)$. \label{footupdown}} in Fig. \ref{fig1a}.
The self and mutual statistical angles can be extracted from the topological theory \eqref{EFT} as
\beq
\theta_l=\pi l^{\text{T}}\cdot K^{-1} \cdot l, \qquad \theta_{l, l'}=2\pi l^{\text{T}}\cdot K^{-1} \cdot l'.
\eeq
Thus, we find that $e$ and $m$ are bosons and $\epsilon$ is a fermion for the $s$-wave superconductor. In addition, any mutual braiding gives the statistical angle $\pi$. 

Now we consider the symmetries: 
In a generic effective theory \eqref{EFT} the charge $Q_A$ of an excitation characterized by the $l$-vector is given by
\beq \label{charge}
Q^A_l=t_A^{T}\cdot K^{-1}\cdot l.
\eeq
For the $s$-wave state, it follows from the derivation in Sec. \ref{microsec} that the $t$-vectors for the spin and magnetic flux symmetry are
$
t^T_s=(2,0,1,-1)$, $t^T_v=(1,0,0,0)
$.
This fixes the spin and vortex charges of $m$, $\epsilon$ and $e$ to be
\beq \label{charges}
\begin{split}
& Q_s^m=0, \qquad Q_s^\epsilon=1, \qquad  Q_s^e=1,\\
& Q_v^m=1/2, \qquad Q_v^\epsilon=0, \qquad Q_v^e=1/2. 
\end{split}
\eeq

Since we previously defined that in a superconductor, Bogoliubov particles carry \emph{only spin} and vortices carry \emph{only magnetic flux}, we now must identify $\epsilon$ and $m$ with the Bogoliubov quasiparticle and vortex respectively. As expected, the Bogoliubov quasiparticle $\epsilon$ is a fermion, while the vortex $m$ is a boson. Their composite $e$ is a \emph{boson}, which carries the spin and vortex charge. The mutual $\pi$-phase under the braiding of $\epsilon$ around $m$ is consistent with the well-known fact that in a superconductor the Bogoliubov quasiparticle (despite being electrically neutral \cite{Kivelson1990}) accumulates a minus sign upon encircling a vortex \cite{Reznik1989, Goldhaber1991}.

The $s$-wave pairing does not violate two-dimensional parity $P$ ($x\leftrightarrow y$) and time-reversal $T$ ($t\to -t$). As a result, the effective theory \eqref{EFT} should be invariant under these discrete symmetries. We specify the transformation properties of all fields:
 First, under $P$ and $T$ the Bogoliubov quasiparticle currents $j_{\uparrow}$, $j_{\downarrow}$ and the vortex current $j_{\text{v}}$ transform differently. The definitions of the currents lead to the following nontrivial transformations
\beq \label{PTcurrents}
\begin{split}
&P:j^0_{\uparrow}\leftrightarrow j^0_{\downarrow}, \quad j^x_{\uparrow}\leftrightarrow j^y_{\downarrow}, \quad j^{0}_{\text{v}}\to -j^{0}_{\text{v}}, \quad j^x_{\text{v}}\leftrightarrow -j^y_{\text{v}},  \\
&T: j^0_{\uparrow}\leftrightarrow j^0_{\downarrow}, \quad j^i_{\uparrow}\leftrightarrow -j^i_{\downarrow}, \quad j^{0}_{\text{v}}\to -j^{0}_{\text{v}}.
\end{split}
\eeq
Given these, the statistical gauge fields $c^{\uparrow}$ and $c^{\downarrow}$ in Eq. \eqref{topfers} transform as
\beq
\begin{split}
&P:c_0^{\uparrow}\leftrightarrow -c_0^{\downarrow}, \quad c_x^{\uparrow}\leftrightarrow -c_y^{\downarrow},  \\
&T: c_0^{\uparrow}\leftrightarrow -c_0^{\downarrow}, \quad c_i^{\uparrow}\leftrightarrow c_i^{\downarrow},
\end{split}
\eeq
while the statistical gauge field $b$, the vortex source $\mathcal{A}^v$ and the spin source $\mathcal{A}^s$ transform like the vortex current $j_{\text{v}}$ in Eq. \eqref{PTcurrents}.
Finally, under $P$ and $T$ the statistical gauge field $a$ transforms like the electromagnetic gauge potential $A$,
\beq
\begin{split}
&P: a^x \leftrightarrow a^y  \\
&T:  a^i\to -a^i.
\end{split}
\eeq
Using these transformation properties, it is straightforward to check that the effective theory \eqref{EFT} is indeed invariant under $P$ and $T$. 

In the absence of external quasiparticle currents in Eq. \eqref{EFT}, all statistical gauge fields can be integrated out resulting in the Hall response
\beq \label{res}
\mathscr{L}_{\text{res}}=-\frac{1}{4\pi} \underbrace{t_A^{\text{T}}\cdot K^{-1} \cdot t_B}_{\nu_{AB}} \epsilon^{\mu\nu\rho} \mathcal{A}^A_{\mu} \partial_\nu \mathcal{A}^B_{\rho}.
\eeq
A simple calculation leads to $\nu_{AB}=0$. Thus the $s$-wave superconductor reassuringly exhibits no Hall effects, which is consistent with $P$ and $T$ invariance of this state.

Finally, we look at the edge, where there are \emph{two pairs of counter-propagating} chiral modes because the $K$-matrix has two pairs of eigenvalues of the opposite sign. 
Note, however, that one is allowed to add to the edge Lagrangian \eqref{edge} two independent Higgs terms $\cos(2 t_s \cdot \phi)$ and $\cos(2 t_v \cdot \phi)$ which satisfy the conditions Eq. \eqref{gapconds}.  
As a result, the two Higgs terms completely gap out all four edge modes of the $s$-wave superconductor, consistent with the expectation that this state neither has gapless edge modes nor a Hall response.
The above arguments explicitly demonstrate that the low energy physics of the $s$-wave state cannot be completely characterized by the toric code model since the edge of the toric code can be gapped out in two physically distinct ways \cite{Bravyi1998} whereas the edge of an $s$-wave superconductor with U$(1)_s\times$U$(1)_v$ symmetry is gapped by the unique mechanism presented above. 


\subsection{$d+ id$ superconductor} \label{dwave}
In a chiral $d+ id$ superconductor, parity and time-reversal are broken spontaneously, which gives rise to \emph{anyon} self-statistics of excitations. 
In parallel with the $s$-wave case, in Sec. \ref{microsec} we derived the fermionic $K$-matrix for this state
\beq \label{Kd}
K=\left(
\begin{tabular}{ cccc}
  0 & 2 &1 & 1 \\
  2 & 0 &0 & 0  \\
  1 & 0 &1 & 0 \\
  1 & 0 &0 & 1 
\end{tabular}
\right).
\eeq
It describes a chiral state with a chiral central charge $c=k=2$. The $l$-vectors are identical to the $s$-wave case and are illustrated in Fig. \ref{fig1a} (also see footnote \cite{Note11}). As a result,
in the weakly-coupled topological phase of a $d+ id$ superconductor, $e$ and $m$ excitations are semions \cite{Kitaev2006}, i.e., they have the statistical angle $\theta_e=\theta_m=\pi/2$. On the other hand, $\epsilon$ is a fermion and has nontrivial mutual $\pi$ statistics with $e$ and with $m$.

The derivation undertaken in Sec. \ref{microsec} fixed the symmetry $t$-vectors for this state to be 
\beq \label{tvecdid}
t^T_s=(0,0,1,-1), \quad t^T_v=(1,0,0,0).
\eeq
Using Eq. \eqref{charge}, we find that the spin and vortex charges of the excitations equal
\beq \label{chargesd}
\begin{split}
& Q_s^m=0, \qquad Q_s^\epsilon=1, \qquad  Q_s^e=1\\
& Q_v^m=1/2 \qquad Q_v^\epsilon=0, \qquad Q_v^e=1/2. 
\end{split}
\eeq
which are, in fact, identical to those for the $s$-wave case given in Eq. \eqref{charges}. As a result, in this case $\epsilon$ and $m$ will be still identified with the Bogoliubov quasiparticle and vortex, respectively. In contrast to the $s$-wave superconductor, the vortex here is a semion. In fact, the semion statistics of the vortex can be extracted from the Berry phase accumulated under exchange of two identical vortices in a $d+id$ superconductor, which can be computed by a simple generalization of the computation done for a $p+ip$ superconductor in \cite{Ariad2015}. Alternatively, the semion phase follows from a qualitative argument presented in \cite{Bernevig2015} that views a $d\pm id$ superconductor as a stack of four spinless $p\pm ip$ layers.

The effective theory \eqref{EFT} with the $K$-matrix \eqref{Kd} is $PT$ invariant, but breaks separately $P$ and $T$ symmetries. Consequently, we find a nontrivial Hall response 
\beq \label{nu}
\nu= \Big(
\begin{tabular}{ cc}
  $2$ & 0 \\
  0 & $0$  \\
\end{tabular}
\Big).
\eeq
with $\nu$ defined in Eq. \eqref{res}.
We thus showed here that in the topological (weakly-coupled) phase a chiral $d$-wave superconductor exhibits the spin Hall effects with $\nu_s=2$, but no vortex ($\nu_v=0$) and mixed spin-vortex Hall ($\nu_{vs}=0$) responses. For the spin part, this reproduces in appropriate units the findings from \cite{Senthil1999, Read2000}. In particular, in a $d+id$ paired state a position-dependent external magnetic Zeeman field $B(x, y)$ will give rise to the Hall current of the $z$-component of spin $j_s^i=- \sigma_s\epsilon^{ij}\partial_j B$ with the spin Hall conductivity $\sigma_s= 1/(4\pi)$ \footnote{Restoring $\hbar$, in our convention the unit of the spin charge is equal to $\hbar/2$ and thus $\sigma_s=\nu_{s} (\hbar/2)^2/h= \hbar/(4\pi)$}.  But what is the physical implication of the absence of the vortex Hall effect? Since the density and current of the elementary fermions fix the background values of $\mathcal{B}^v$ and $\mathcal{E}_i^v$ (see Appendix \ref{App1}), a nontrivial vortex Hall effect \emph{would} imply
\beq
\begin{split}
n_v&=\frac{\nu_v}{2\pi} \mathcal{B}^v=\nu_v n_f, \\
j^i_v&=-\frac{\nu_v}{2\pi} \epsilon^{ij}\mathcal{E}_j^v=\nu_v j_f^i,
\end{split}
\eeq
i.e., a linear relation between the densities of the elementary fermions and vortices in the ground state of the $d+id$ superconductor. The fact that we found $\nu_v=0$ demonstrates explicitly that $\emph{zero}$ magnetic field $B$ (an thus zero density of vortices) is generated in the ground state of the chiral $d$-wave superconductor. At first sight it might seem surprising that the unbroken magnetic flux symmetry has a vanishing Hall response in the $d+id$ paired state that breaks spontaneously parity and time-reversal. In Sec. \ref{noHall} we present a general argument which independently supports this finding.

Consider now the edge of a chiral $d$-wave superconductor \eqref{Kd}, where \emph{two co-propagating} chiral bosons appear together with a \emph{pair of counter-propagating} chiral modes. For the purpose of the upcoming discussion, it is convenient to transform the Chern-Simons theory into a GL$(4,\mathbb{Z})$ equivalent form $(K,l,t)\to(\tilde K, \tilde l, \tilde t)$, discussed in detail in Appendix \ref{App2}. In this formulation, the $K$-matrix is \emph{block-diagonal}
\beq \label{Ktd}
\tilde K_B=\left(
\begin{tabular}{ cccc}
  -2 & 2 &0 & 0 \\
  2 & 0 &0 & 0  \\
  0 & 0 &1 & 0 \\
  0 & 0 &0 & 1 
\end{tabular}
\right),
\eeq
the $l$-vectors are
\beq
\begin{split}
&\tilde l_m^T=(0,1,0,0)=l^T_m, \\
&\tilde l_{\epsilon\uparrow}^T=(-1,0,1,0), \quad \tilde l_{\epsilon\downarrow}^T=(-1,0,0,1),  
\end{split}
\eeq
and the $t$-vectors are unchanged from Eq. \eqref{tvecdid}: $\tilde t_s=t_s$, $\tilde t_v=t_v$. In this basis, the counter-propagating modes $\tilde \phi^1$ and $\tilde \phi^2$ are gapped out by the Higgs term $\cos(2 \tilde t_v \cdot \tilde\phi)$ that fulfils the conditions \eqref{gapconds} introduced at the end of Sec. \ref{WZ}. On the other hand,  due to quantization of the thermal Hall conductance \cite{Kane1997, Kitaev2006}, the remaining two co-propagating chiral states $\tilde \phi^3$ and $\tilde \phi^4$ cannot be gapped by edge interactions or disorder. These chiral states are neutral under the magnetic flux U$(1)_v$ symmetry. This implies that there is no U$(1)_v$ gauge anomaly at the edge, consistent with the vanishing vortex Hall effect found above (see also Sec. \ref{noHall}). In contrast, the spin Hall effect with $\nu_s=2$ implies that the edge theory must have a spin gauge anomaly. In fact, the current associated with the spin chiral boson $\tilde l_{s} \cdot \tilde \phi$ with
\beq
l_{s}^T=\tilde l_{\epsilon\uparrow}^T-\tilde l_{\epsilon\downarrow}^T=(0,0,1,-1)
\eeq
realizes the U$(1)_2$ Kac-Moody affine algebra, where the subscript denotes the level. From a standard argument \cite{Ginsparg1988}, the current affine algebra U$(1)_2$ can be actually extended to a larger affine algebra SU$(2)_1$. The spin sector is thus described by the chiral SU$(2)_1$ Wess-Zumino-Witten edge theory \cite{Read2000} that has the spin gauge anomaly. The complete bulk plus edge theory is of course consistent because the edge anomaly is canceled by the inflow of the bulk spin Hall current.

Curiously, the current associated with the orthogonal combination
$
l_{tot}^T=\tilde l_{\epsilon\uparrow}^T+\tilde l_{\epsilon\downarrow}^T=(-2,0,1,1)
$
realizes another copy of the U$(1)_2\to$ SU$(2)_1$ Kac-Moody affine algebra. The above arguments thus suggest that the edge theory of the chiral $d$-wave superconductor has an extra SU$(2)_{tot}$ symmetry. As discussed in \cite{Read2000}, the $SU(2)_{tot}$ is emergent and does not have a microscopic origin. In a $d\pm id$ superconductor, with the spin and particle-hole symmetries, the velocities of the two co-propagating chiral modes are necessarily \emph{equal}. Therefore the edge supports an anomalous SO$(4)\cong$ SU$(2)_s\otimes $SU$(2)_{tot}$ symmetry which rotates four edge Majorana modes; for more details, see \cite{Read2000}. In the following subsection, we show how this discussion generalizes to superconductors paired in higher partial even-waves.


\subsection{Higher partial waves and the 16-fold way} \label{higher}
So far, we have analyzed Chern-Simons theories for the specific cases of conventional $s$-wave and chiral $d$-wave superconductors, the latter being the simplest example of a spin-singlet chiral state. At this point we extend the discussion to include all chiral spin-singlet superconductors paired in partial even-waves. 

\begin{figure}[ht]
\begin{center}
\includegraphics[height=0.25\textwidth]{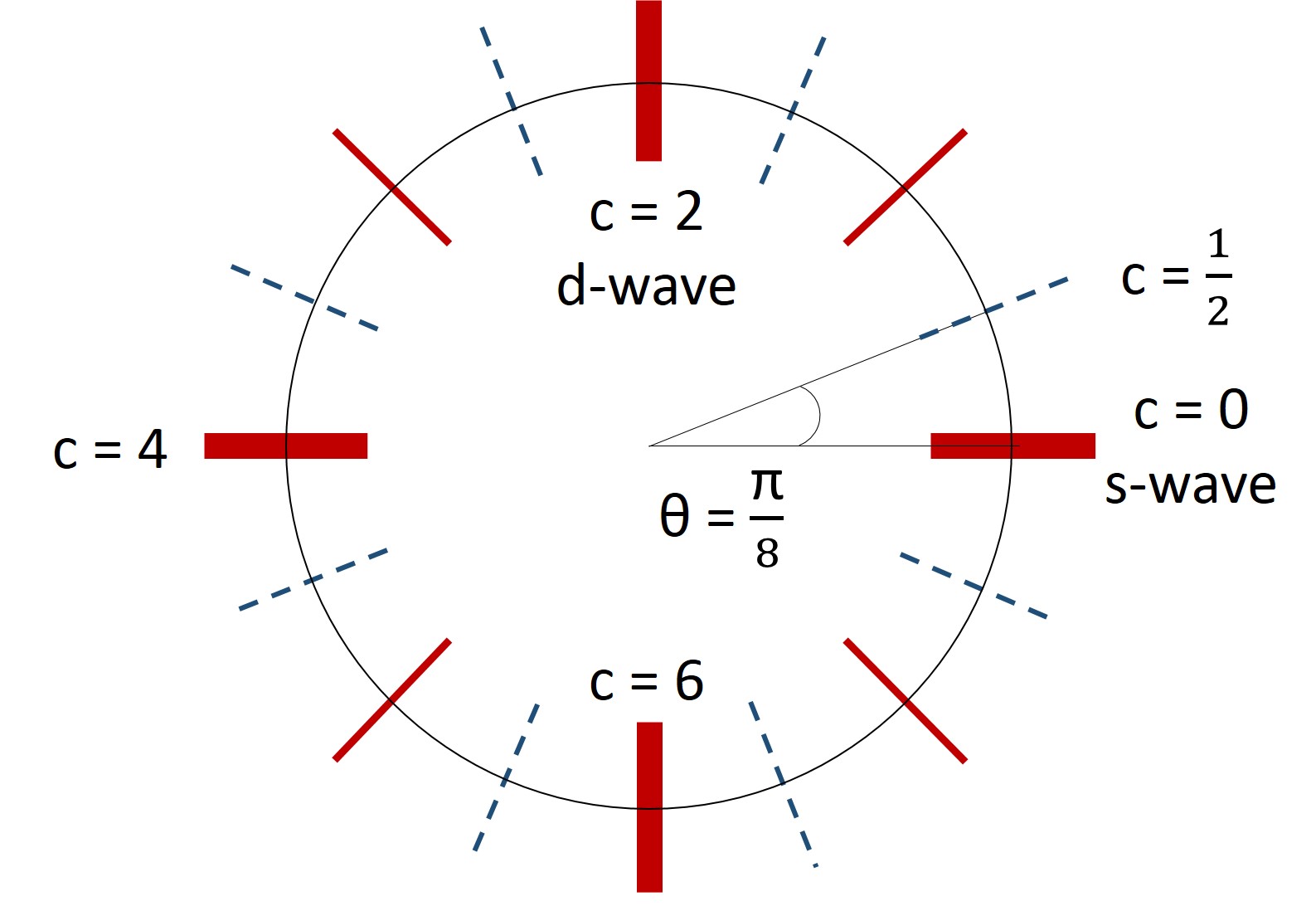}
\caption{(Color online) Diagrammatic representation of the 16-fold way. The abelian states considered in this paper are represented by thick lines and the other abelian states by thin lines. Dashed lines indicate the non-abelian states.  For each state, the angle $\theta$ equals the exchange angle $\theta_{e}=\theta_{m}$.}\label{fig2a}
\end{center}
\end{figure}

In \cite{Kitaev2006}, Kitaev demonstrated that chiral superconductors have a $\mathbb{Z}_{16}$ \emph{bulk} classification. Using either the language of axiomatic topological field theory or by considering stacked $p + $i$p$ layers \cite{Bernevig2015}, it can be shown that the statistical angle acquired upon exchanging two $e$-particles (or two $m$-particles) has a 16-fold periodicity. That is, given a chiral superconductor with Chern number $\nu$ that hosts $\nu$ Majorana chiral modes at the edge, the exchange angle of excitations \cite{Kitaev2006}
\beq \label{16stat}
\theta_e = \theta_m = 2 \pi \nu/16, \qquad \theta_\epsilon=\pi.
\eeq
Since every Majorana mode contributes a half unit of the central charge, the total chiral central charge $c=\nu/2$. Thus, systems with $c = 0$ and $c = 8$ have identical bulk anyonic excitations but have different edge theories manifested in different thermal Hall conductivities. The 16-fold way is diagrammatically depicted in Fig.~\ref{fig2a}, where the angle $\theta$ represents $\theta_{e}$ for the different states. The states with half-integer central charge (represented by dashed lines) are \emph{non-abelian} whereas those with integer central charge (represented by solid lines) are \emph{abelian}. Among the latter, the thick lines indicate states with $c \in 2\mathbb{Z}$, which are realized by the spin-singlet superconductors studied in this paper. For these superconductors, the chiral central charge $c$ equals the chirality parameter $k$.

From Sec. \ref{microsec}, it follows that the weakly-paired chiral superconductor paired in even $k^{\text{th}}$ partial wave has the $(k+2)\times(k+2)$ $K$-matrix
\beq \label{Kmat}
K=\left(
\begin{array}{cc|c}
 0 & 2 & 1_k  \\
 2 & 0 & 0_k  \\ \hline
 1_k^T & 0^T_k & \mathbb{1}_{k\times k}
 \end{array}
\right),
\eeq
where we introduced the notation $x_k=(\underbrace{x,x,\dots,x}_{k \text{ times}})$. In fact, this $K$-matrix has previously appeared in a somewhat different context in \cite{You2015}.
The $l$-vectors of the excitations are
\beq \label{lhigher}
\begin{split}
l^T_m&=(0,1,0_k), \\
l^T_\epsilon&=(0,0,1,0_{k-1}).
\end{split}
\eeq
It is straightforward to demonstrate that these spin-singlet states $k\in 2 \mathbb{Z}/\{0\}$ fall into the 16-fold way since they have the chiral central charge $c=k$ and lead to the statistics \eqref{16stat}. Interestingly, the states where $c$ is an odd integer form a different class of \emph{abelian} states which perhaps describe certain phases of spin-triplet superconductors.

The charges of excitations are given by Eq. \eqref{chargesd}, i.e., they are identical to the $s$-wave and $d+id$ cases. These states are fixed by the symmetry $t$-vectors, which in this case are given by
\beq \label{thigher}
\begin{split}
t^T_s&=(0,0,\pm1_k), \\
t^T_v&=(1,0,0_k),
\end{split}
\eeq
where $\pm1_k=(\underbrace{1,-1,\dots,1,-1}_{k/2 \text{ times}})$.
We thus again find the Bogolibuov quasiparticle to be the fermion $\epsilon$, while the vortex is the anyon $m$ that has the statistical angle $\theta_m=\pi k/8$. An explicit calculation gives the spin Hall coefficient $\nu_s=k$ and the vanishing vortex ($\nu_v=0$) and mixed vortex-spin ($\nu_{vs}=0$) coefficients. This result is a natural generalization of Eq. \eqref{nu} found for the $d+id$ state.

\begin{figure}[ht]
\begin{center}
\includegraphics[width=0.5\textwidth]{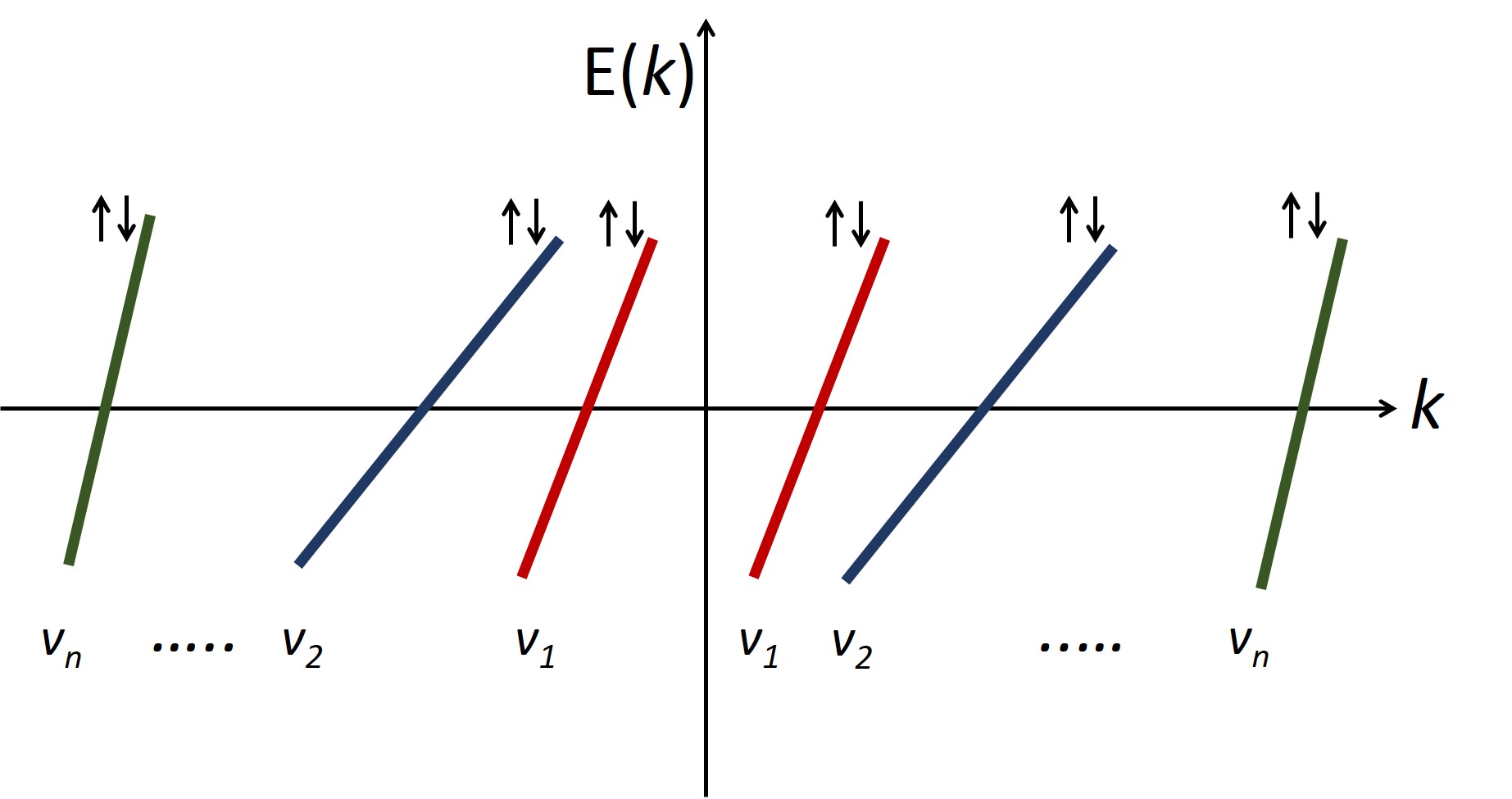}
\caption{A chiral superconductor with chirality $k \in 2\mathbb{Z}$ hosts $2k$ Majorana modes at the edge. As indicated by the $(\uparrow \downarrow)$ arrows, each mode is doubly degenerate due to spin symmetry and modes with the same color are related by particle-hole symmetry. Thus, the modes within a quartet have the same velocity $v_i$, while generically different quartets will have different velocities.}\label{fig3a}
\end{center}
\end{figure}

One can better understand the edge theory of chiral paired states by casting the $K$-matrices \eqref{Kmat} into two different GL$(k+2,\mathbb{Z})$-equivalent block-diagonal forms $\tilde K_B$ and $\tilde K_C$ (see Appendix \ref{App2} for details). Generalizing the discussion in Sec. \ref{dwave}, from the form
\beq
\tilde K_B=\left(
\begin{array}{cc}
 -k & 2   \\
 2 & 0   \\   
 \end{array}
\right) \oplus \mathbb{1}_{k\times k},
\eeq
one finds that \emph{a pair} of counter-propagating edge modes is gapped out by the allowed Higgs term $\cos(2 \tilde t_v \cdot \phi)$. Similar to the $d+id$ case, the remaining gapless chiral theory is neutral under the magnetic flux U$(1)_v$ symmetry.
On the other hand, symmetries of the edge appear naturally from  the Cartan block-diagonal form
\beq \label{KCartan}
\tilde K_C= A^{\text{SO}(2k)}_{k\times k} \oplus \left(
\begin{array}{cc}
 1 & 0   \\
 0 & -1   \\   
 \end{array}
\right),
\eeq
where $A^{\text{SO}(2k)}_{k\times k}$ is the \emph{Cartan matrix} of the Lie algebra SO$(2k)$ and is defined in Eq. \eqref{Cartan}.
For example, for the $d+id$ superconductor ($k=2$), one finds $\tilde K_C=\left(
\begin{array}{cc}
 2 & 0   \\
 0 & 2   \\   
 \end{array}
\right)  \oplus \left(
\begin{array}{cc}
 1 & 0   \\
 0 & -1   \\   
 \end{array}
\right),
$
with the SO$(4)\cong$ SU$(2)_s\otimes $SU$(2)_{tot}$ symmetry, discussed in Sec. \ref{dwave}, manifest.
In a similar fashion, following \cite{Ginsparg1988}, we can show that the current operators arising from the Cartan block satisfy $SO(2 k)_1$ Kac-Moody algebra.
This seems to suggest that the chiral edge theory might have an internal SO$(2k)$ symmetry associated with rotations of the $2k$ multiplet of Majorana modes. 
Generically, however, we expect this symmetry to be \emph{broken} by the velocity matrix. 
As illustrated in Fig. \ref{fig3a}, in a chiral superconductor of chirality $k$, $2k$ Majorana edge modes split into $n=k/2$ quartets. The modes within each quartet are related by the spin and particle-hole symmetry, and thus have the same velocity. However, nothing prevents different quartets from having different velocities. As a result the nature of the residual symmetry at the edge of a chiral superconductor depends on the velocity matrix and hence on the microscopic details.

From the Cartan form \eqref{KCartan} of the $K$-matrices, it is clear that the fermionic block, $\sigma^z$, is topologically trivial, i.e., it does not effect the ground state degeneracy, chirality, and statistics of the system. In other words, in the absence of symmetries these $K$-matrices are stably equivalent \cite{Cano2014} to the Cartan matrices $A^{\text{SO}(2k)}$ \footnote{Incidentally, these purely bosonic Cartan $K$-matrices have also been proposed for describing $s$-wave superconductors strongly proximity-coupled to topological Chern insulators in \cite{Teo2015}.}. 
However, in the presence of coupling to the vortex source $\mathcal{A}^v$, the fermionic block is essential because the vortex symmetry vector $t_v$ (in this basis) has non-zero elements in this sector.

\section{Absence of vortex Hall effect in superconductors} \label{noHall}
In the previous section we found for all superconductors that the vortex Hall effect for the magnetic flux symmetry is zero. While for the $s$-wave state this is a completely expected result, for chiral spin-singlet states it might seem surprising since parity and time-reversal are spontaneously broken and one might expect that these unbroken symmetries should exhibit nontrivial Hall responses. Here we will argue that the vortex Hall effect should vanish in \emph{any} superconductor due to the finite mass of the photon field acquired via the Higgs mechanism.

\begin{figure}[ht]
\begin{center}
\includegraphics[width=0.3\textwidth]{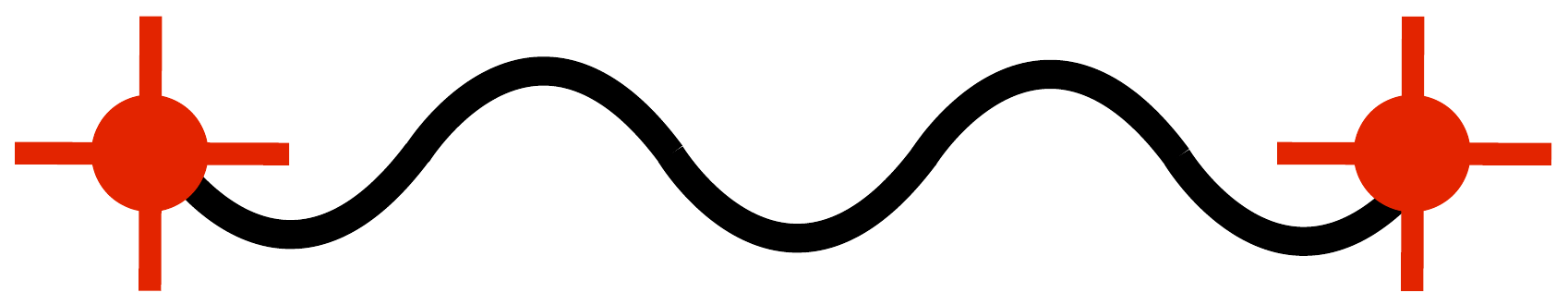}
\caption{Feynman diagram for the kernel $\Gamma^{\mu\nu}(p)$: the bold wavy line denotes the renormalized photon propagator $iD_{\gamma\delta}(p)$, the red cross represents the vertex originating from the last term in Eq. \eqref{SCgen}.}\label{fig4}
\end{center}
\end{figure}

Consider a general two-dimensional superconductor coupled to the magnetic flux source $\mathcal{A}^v$
\beq \label{SCgen}
\mathcal{L}=\mathcal{L}_{\text{SC}}(f,A)-\frac{1}{2\pi} \epsilon^{\mu\nu\rho}\mathcal{A}^v_\mu \partial_\nu A_\rho,
\eeq
where the elementary fermions $f$ undergo Cooper pairing and thus generate a mass to the electromagnetic gauge potential $A$ via the Higgs mechanism. After integrating out $f$ and $A$, the vortex Hall effect can in principle appear from the quadratic contribution to the effective action $\Gamma[\mathcal{A}^v]$. In a translation-invariant system
\beq
\begin{split}
\Gamma^{(2)}[\mathcal{A}^v]&=\frac 1 2 \int d^3x d^3y \mathcal{A}^v_\mu(x) \Gamma^{\mu\nu}(x-y) \mathcal{A}^v_\mu(y) \\
&=\frac 1 2 \int d^3p \mathcal{A}^v_\mu(-p) \Gamma^{\mu\nu}(p) \mathcal{A}^v_\mu(p),
\end{split}
\eeq
where in momentum space the kernel $\Gamma^{\mu\nu}(p)$ is given by
\beq
\Gamma^{\mu\nu}(p) \sim \epsilon^{\mu\alpha\gamma}p_\alpha D_{\gamma \delta} (p) \epsilon^{\nu \beta \delta} p_{\beta}
\eeq
with $iD_{\gamma\delta}(p)$ a fully renormalized photon propagator.
The kernel is illustrated in Fig. \ref{fig4} as a Feynman diagram.

Let us start in the $s$-wave phase which obviously has a vanishing vortex Hall coefficient $\nu_v=0$. As in Sec. \ref{microsec}, by tuning the Dirac masses of the Bogoliubov quasiparticles we can enter into the chiral $d+id$ phase (see Fig. \ref{fig3}). Importantly, during this process the photon propagator \emph{always} remains gapped due to the Higgs mechanism. Since $\nu_v$ can only change at a point where the photon gap closes, this shows that as we enter the $d+id$ phase $\nu_v$ remains zero. This result should be contrasted to the spin Hall effect which arises from the \emph{fermionic} one-loop diagram \cite{Senthil1999, Read2000}. It is clear from Fig. \ref{fig3} that after starting in the $s$-wave state the fermionic gap must close as one enters the $d+id$ phase, allowing the spin Hall coefficient to jump from $\nu_s=0$ ($s$-wave) to $\nu_s=2$ ($d+id$).

Clearly the argument above can be generalized to other chiral superconductors that can be obtained by deformations of a time-reversal invariant superconducting state. In summary, as long as the Higgs mechanism produces a finite gap for the photon field (along the deformation trajectory), the Hall coefficient for the magnetic flux symmetry should \emph{vanish}. This is in a stark contrast to the standard probes of time-reversal breaking, such as the polar Kerr, spin Hall and Nernst/Ettingshausen effects that should give non-vanishing signals in chiral superconducting states.


\section{Open questions and outlook}\label{disc}
In this work we have developed a topological framework for the low-energy physics of two-dimensional spin-singlet superconductors. Here we discuss some open questions that go beyond the scope of this paper and are left for future studies.

As emphasized throughout the paper, in this work the electromagnetic field is strictly confined to \emph{two} spatial dimensions. On the other hand, in thin film superconductors the \emph{mixed-dimensional} problem of two-dimensional paired fermions interacting with a three-dimensional electromagnetic field is realized most naturally. For this reason, this system deserves to be studied in detail. It also might fall into the class of quasi-topological phases introduced by Bonderson and Nayak in \cite{Bonderson2013}.

Additionally, here we have considered a theory that allows topological defects (vortices) of the matter field, but not topological defects (magnetic monopoles) in the electromagnetic (non-compact) sector. It might be interesting to investigate a two-dimensional superconductor with a \emph{compact} electromagnetism; in this version of the theory, magnetic monopoles are allowed and appear in the form of spacetime instantons \cite{Polyakovbook}. As a result, the magnetic flux symmetry is lost because a pair of vortices can instantly disappear into a monopole. This rich problem naturally arises in the physics of spin liquids and was investigated in the seminal paper of Fradkin and Shenker \cite{Fradkin1979}. In the future, it would be interesting to extend our work to this model and to investigate the interplay of the unbroken $SU(2)_s$ spin symmetry and topological order in it. 

We have demonstrated in this paper that all superconductors have a vanishing vortex Hall response due to the finite gap of the photon field that arises from the Higgs mechanism. As long as this gap can be closed, however, it seems possible that one can enter a distinct phase of matter with $\nu_v\ne 0$. In contrast to a superconductor, this phase is characterized by a finite flux of magnetic field in the ground state, corresponding to a dense collection of vortices. It will hence be of interest to find ways of closing the photon gap in our problem.

Moreover, the realization of the magnetic flux symmetry might be subtle in the problem studied here. For a bosonic toric code model, for instance, it is known that it is \emph{impossible} to realize an internal global $U(1)$ symmetry with the charges $Q_e=Q_m=1/2$ in a purely two-dimensional world. In fact, this realization of a symmetry can only appear on the surface of a three-dimensional system. This is known as the statistical anomaly and, for bosonic systems, is discussed in detail in \cite{Metlitski2013, Chong2013}. We believe that the fate of this anomaly in two-dimensional fermionic superconductors deserves future investigation.

Recently, it has also become clear that in a weakly paired $d+id$ superfluid (and also in chiral superfuids paired in higher partial waves) not all fermions in the ground state are paired in the presence of an edge \cite{Tada2015}. These unpaired fermions are localized close to the edge and carry a mass current that partially compensates the angular momentum $L=l N/2$ carried by the chiral Cooper pairs. This current is \emph{non-universal} since it depends on the structure of the edge. It would be interesting to investigate the nature of this current in the presence of a dynamical electromagnetism. 

A limitation of the $K$-matrix formalism is that it allows us to incorporate only the coupling to the abelian subgroup of the spin symmetry. Currently, the Chern-Simons field theory does not allow coupling to the full nonabelian spin symmetry because we do not know how to fractionalize representations of nonabelian groups. Using an alternative formalism that allows coupling to all non-abelian SU$(2)_s$ sources could thus generalize our work. In addition, since any physical system breaks the spin symmetry due to \emph{spin-orbit coupling}, it would be useful to study the effects of (weakly) breaking this symmetry.

Finally, it is known that two-dimensional chiral superconductors and superfluids exhibit a non-vanishing ``shift'' and Hall viscosity \cite{Read2009, Read2011,Hoyos2013}. Following the seminal work of Wen and Zee \cite{Wen1992}, it should be possible to account for these phenomena by coupling the spin connection to the topological Chern-Simons theories developed here.

\section*{Acknowledgments:}
We gratefully acknowledge discussions with Leonid Glazman, Michael Hermele, Alexei Kitaev, Max Metlitski, Rahul Nandkishore, Chong Wang, and participants of the workshop ``Geometry of Quantum states in Condensed
Matter Systems'' at Simons Center for Geometry and Physics.
SM gratefully acknowledges support from the Simons Center for Geometry and Physics, Stony Brook University at which some of the research for this paper was performed.
SM and LR acknowledge support by the NSF grants DMR-1001240, 
PHY-1125915, and by the Simons Investigator award from the Simons 
Foundation. 
AP and VG acknowledge support by the NSF grants DMR-1205303 and PHY-1211915.
The work of SM is supported by the Emmy Noether Programme of German Research Foundation (DFG) under grant No. MO 3013/1-1.
We thank the KITP for its
hospitality during our stay as part of the ``Universality in Few-Body Systems"
(SM), ``Synthetic Quantum Matter'' (LR) and sabbatical (LR) programs, when part
of this work was completed.

\appendix
\section{Relation between ``magnetic field'' $\mathcal{B}^v$ and fermion density $n_f$}\label{App1}
Here we demonstrate that in a superconductor the background vortex ``magnetic field'' $\mathcal{B}^v=\epsilon^{ij}\partial_i \mathcal{A}^v_j$ is fixed by the density of the elementary fermions $n_f$. Indeed, in a superconductor the Lagrangian of a single vortex located at position $X^i$ is given by \cite{Cooper2008}
\beq
L_v=- \frac{\pi}{2} n_f \epsilon_{ij}X^i \dot{X}^j.
\eeq
Alternatively, we can rewrite the same Lagrangian as a minimal coupling of a vortex current to its gauge potential $L_v=-q_v \mathcal{A}^{v}_j \dot X^j$ with $q_v=1/2$ because in a superconductor the vortex carries only a half of a unit flux quantum of the magnetic field. As a result, we find that the density of elementary fermions fixes the background value of $\mathcal{B}^v$ to be $\mathcal{B}^v=2\pi n_f$. At the same time the fermion current $j^i_f$ fixes the background value of the ``electric field'' $\mathcal{E}_j^v$ via the relation 
$ j_f^i=-\frac{1}{2\pi} \epsilon^{ij}\mathcal{E}_j^v$.
\section{Block-diagonal forms of $K$-matrices of spin-singlet chiral states} \label{App2}
It is well-known that different $K$-matrices can represent the same topological state in the Chern-Simons theory \eqref{EFT}. Indeed, one is allowed to relabel the statistical gauge fields $a^I$
\beq
a^I\to \tilde a^I= X_{IJ} a^J,
\eeq
where $X\in GL(N,\mathbb{Z})$ is a $N\times N$ matrix of integers with the determinant $\pm 1$. Under such a transformation
\beq
\begin{split}
&K\to\tilde K= X^T K X, \\
&l\to \tilde l=X^T l, \\
&t\to \tilde t=X^T t
\end{split}
\eeq
and the theories defined by $(K,l,t)$ and $(\tilde K,\tilde l,\tilde t)$ are \emph{equivalent}.

We demonstrate that the $K$-matrix of a chiral spin-singlet superconductor paired in the (even) $k^{\text{th}}$ partial wave can be put into \emph{block-diagonal} forms. The two distinct block-diagonal forms used in the main text are highlighted here. We start from the $K$-matrix, $l$-vectors, and $t$-vectors that were derived in Sec. \ref{higher} and are given by Eqs. \eqref{Kmat}, \eqref{lhigher}, and \eqref{thigher} respectively.

If one chooses now the $X$-matrix
\beq \label{Xmat}
X=\left(
\begin{array}{cc|c}
 1 & 0 & 0_k  \\
 0 & 1 & 0_k  \\ \hline
 -1_k^T & 0^T_k & \mathbb{1}_{k\times k}
 \end{array}
\right),
\eeq
one finds
\beq \label{Kt1}
\tilde K_B=\left(
\begin{array}{cc}
 -k & 2   \\
 2 & 0   \\   
 \end{array}
\right) \oplus \mathbb{1}_{k\times k},
\eeq
\beq \label{lhigherapp}
\begin{split}
\tilde l^T_m&=(0,1,0,\dots,0)=l^T_m, \\
\tilde l^T_\epsilon&=(-1,0,1,0,\dots,0),
\end{split}
\eeq
\beq \label{thigherapp}
\begin{split}
\tilde t^T_s&=(0,0,1,-1,\dots, 1, -1)=t^T_s, \\
\tilde t^T_v&=(1,0,\dots,0)=t^T_v.
\end{split}
\eeq
In Eq. \eqref{Kt1} the topological order and chirality is encoded in the first and second factor, respectively.

Alternatively, one can find a $GL(N,\mathbb{Z})$ transformation that transforms the $K$-matrix \eqref{Kmat} into the Cartan block-diagonal form
\beq \label{Kt2}
\tilde K_C= A^{\text{SO}(2k)}_{k\times k} \oplus \left(
\begin{array}{cc}
 1 & 0   \\
 0 & -1   \\   
 \end{array}
\right),
\eeq
where we introduced the \emph{Cartan matrix} of of the Lie algebra SO$(2k)$
\beq \label{Cartan}
A^{\text{SO}(2k)}_{k\times k}=\left(
\begin{array}{cccccc}
 2 & -1 & 0 & 0 & \dots & 0 \\
 -1 & 2 & \ddots & \ddots & \ddots & 0 \\
 0 & \ddots & 2 & -1 & 0 & 0 \\
 0 & \ddots & -1 & 2 & -1 & -1 \\
 \vdots & \ddots & 0 & -1 & 2 & 0 \\
 0 & 0 & 0 & -1 & 0 & 2 \\
\end{array}
\right).
\eeq
In the form \eqref{Kt2} both the topological order and chirality are encoded in the first factor, while the second factor is topologically trivial. The $X$ matrix for the $d+id$ ($k=2$) case is 
\beq
X_{k=2}=\left(
\begin{array}{cccc}
 -1 & -1 & 0 & -1 \\
 -1 & 0 & 0 & 0 \\
 1 & -1 & 0 & 0 \\
 1 & 1 & 1 & 1 \\
\end{array}
\right).
\eeq
$X$ matrices for higher partial even-waves can also be found. For $k= 4$, $k=6$, and $k=8$, for instance, we find
\beq
X_{k=4}=\left(
\begin{array}{cccccc}
 -1 & 0 & 0 & 0 & 0 & -1 \\
 -1 & 0 & 0 & 1 & 0 & 0 \\
 0 & 0 & 1 & -1 & 0 & 0 \\
 0 & 1 & -1 & -1 & 0 & 0 \\
 1 & -1 & 0 & 0 & 0 & 0 \\
 1 & 0 & 0 & 0 & 1 & 1 \\
\end{array}
\right),
\eeq
\beq
X_{k=6}=\left(
\begin{array}{cccccccc}
 -1 & 0 & 0 & 0 & 0 & 0 & 0 & -1 \\
 -1 & 0 & 0 & 0 & 0 & 1 & 0 & 0 \\
 0 & 0 & 0 & 0 & 1 & -1 & 0 & 0 \\
 0 & 0 & 0 & 1 & -1 & -1 & 0 & 0 \\
 0 & 0 & 1 & -1 & 0 & 0 & 0 & 0 \\
 0 & 1 & -1 & 0 & 0 & 0 & 0 & 0 \\
 1 & -1 & 0 & 0 & 0 & 0 & 0 & 0 \\
 1 & 0 & 0 & 0 & 0 & 0 & 1 & 1 \\
\end{array}
\right),
\eeq
\beq
X_{k=8}=\left(
\begin{array}{cccccccccc}
 -1 & 0 & 0 & 0 & 0 & 0 & 0 & 0 & 0 & -1 \\
 -1 & 0 & 0 & 0 & 0 & 0 & 0 & 1 & 0 & 0 \\
 0 & 0 & 0 & 0 & 0 & 0 & 1 & -1 & 0 & 0 \\
 0 & 0 & 0 & 0 & 0 & 1 & -1 & -1 & 0 & 0 \\
 0 & 0 & 0 & 0 & 1 & -1 & 0 & 0 & 0 & 0 \\
 0 & 0 & 0 & 1 & -1 & 0 & 0 & 0 & 0 & 0 \\
 0 & 0 & 1 & -1 & 0 & 0 & 0 & 0 & 0 & 0 \\
 0 & 1 & -1 & 0 & 0 & 0 & 0 & 0 & 0 & 0 \\
 1 & -1 & 0 & 0 & 0 & 0 & 0 & 0 & 0 & 0 \\
 1 & 0 & 0 & 0 & 0 & 0 & 0 & 0 & 1 & 1 \\
\end{array}
\right).
\eeq
While these $X$ matrices are not unique, the ones presented here easily generalize to higher $k$.

\newpage

\bibliography{library}

\begin{thebibliography}{74}%
\makeatletter
\providecommand \@ifxundefined [1]{%
 \@ifx{#1\undefined}
}%
\providecommand \@ifnum [1]{%
 \ifnum #1\expandafter \@firstoftwo
 \else \expandafter \@secondoftwo
 \fi
}%
\providecommand \@ifx [1]{%
 \ifx #1\expandafter \@firstoftwo
 \else \expandafter \@secondoftwo
 \fi
}%
\providecommand \natexlab [1]{#1}%
\providecommand \enquote  [1]{``#1''}%
\providecommand \bibnamefont  [1]{#1}%
\providecommand \bibfnamefont [1]{#1}%
\providecommand \citenamefont [1]{#1}%
\providecommand \href@noop [0]{\@secondoftwo}%
\providecommand \href [0]{\begingroup \@sanitize@url \@href}%
\providecommand \@href[1]{\@@startlink{#1}\@@href}%
\providecommand \@@href[1]{\endgroup#1\@@endlink}%
\providecommand \@sanitize@url [0]{\catcode `\\12\catcode `\$12\catcode
  `\&12\catcode `\#12\catcode `\^12\catcode `\_12\catcode `\%12\relax}%
\providecommand \@@startlink[1]{}%
\providecommand \@@endlink[0]{}%
\providecommand \url  [0]{\begingroup\@sanitize@url \@url }%
\providecommand \@url [1]{\endgroup\@href {#1}{\urlprefix }}%
\providecommand \urlprefix  [0]{URL }%
\providecommand \Eprint [0]{\href }%
\providecommand \doibase [0]{http://dx.doi.org/}%
\providecommand \selectlanguage [0]{\@gobble}%
\providecommand \bibinfo  [0]{\@secondoftwo}%
\providecommand \bibfield  [0]{\@secondoftwo}%
\providecommand \translation [1]{[#1]}%
\providecommand \BibitemOpen [0]{}%
\providecommand \bibitemStop [0]{}%
\providecommand \bibitemNoStop [0]{.\EOS\space}%
\providecommand \EOS [0]{\spacefactor3000\relax}%
\providecommand \BibitemShut  [1]{\csname bibitem#1\endcsname}%
\let\auto@bib@innerbib\@empty
\bibitem [{\citenamefont {Wen}(2004)}]{wenbook}%
  \BibitemOpen
  \bibfield  {author} {\bibinfo {author} {\bibfnamefont {X.}~\bibnamefont
  {Wen}},\ }\href {https://books.google.com/books?id=llnlrfdR4YgC} {\emph
  {\bibinfo {title} {Quantum Field Theory of Many-Body Systems}}},\ Oxford
  Graduate Texts\ (\bibinfo  {publisher} {OUP Oxford},\ \bibinfo {year}
  {2004})\BibitemShut {NoStop}%
\bibitem [{\citenamefont {Fradkin}(2013)}]{Fradkin2013}%
  \BibitemOpen
  \bibfield  {author} {\bibinfo {author} {\bibfnamefont {E.}~\bibnamefont
  {Fradkin}},\ }\href
  {http://books.google.com/books?hl=en\&lr=\&id=x7\_6MX4ye\_wC\&oi=fnd\&pg=PR11\&dq=Field+Theories+of+Condensed+Matter+Physics\&ots=OTMzLv0\_tI\&sig=iT7am\_ANJcpca8PXm7hBPm7Dl1E}
  {\emph {\bibinfo {title} {{Field Theories of Condensed Matter Physics}}}}\
  (\bibinfo  {publisher} {Cambridge University Press},\ \bibinfo {year}
  {2013})\BibitemShut {NoStop}%
\bibitem [{\citenamefont {{Zeng}}\ \emph {et~al.}()\citenamefont {{Zeng}},
  \citenamefont {{Chen}}, \citenamefont {{Zhou}},\ and\ \citenamefont
  {{Wen}}}]{Zeng2015}%
  \BibitemOpen
  \bibfield  {author} {\bibinfo {author} {\bibfnamefont {B.}~\bibnamefont
  {{Zeng}}}, \bibinfo {author} {\bibfnamefont {X.}~\bibnamefont {{Chen}}},
  \bibinfo {author} {\bibfnamefont {D.-L.}\ \bibnamefont {{Zhou}}}, \ and\
  \bibinfo {author} {\bibfnamefont {X.-G.}\ \bibnamefont {{Wen}}},\ }\href@noop
  {} {\ }\Eprint {http://arxiv.org/abs/1508.02595} {arXiv:1508.02595
  [cond-mat.str-el]} \BibitemShut {NoStop}%
\bibitem [{\citenamefont {Kivelson}\ and\ \citenamefont
  {Rokhsar}(1990)}]{Kivelson1990}%
  \BibitemOpen
  \bibfield  {author} {\bibinfo {author} {\bibfnamefont {S.~A.}\ \bibnamefont
  {Kivelson}}\ and\ \bibinfo {author} {\bibfnamefont {D.~S.}\ \bibnamefont
  {Rokhsar}},\ }\href {\doibase 10.1103/PhysRevB.41.11693} {\bibfield
  {journal} {\bibinfo  {journal} {Phys. Rev. B}\ }\textbf {\bibinfo {volume}
  {41}},\ \bibinfo {pages} {11693} (\bibinfo {year} {1990})}\BibitemShut
  {NoStop}%
\bibitem [{\citenamefont {Wen}(1991)}]{Wen1991a}%
  \BibitemOpen
  \bibfield  {author} {\bibinfo {author} {\bibfnamefont {X.-G.}\ \bibnamefont
  {Wen}},\ }\href {\doibase 10.1142/S0217979291001541} {\bibfield  {journal}
  {\bibinfo  {journal} {Int. J. Mod. Phys. B}\ }\textbf {\bibinfo {volume}
  {05}},\ \bibinfo {pages} {1641} (\bibinfo {year} {1991})}\BibitemShut
  {NoStop}%
\bibitem [{\citenamefont {Balents}\ \emph {et~al.}(1998)\citenamefont
  {Balents}, \citenamefont {Fisher},\ and\ \citenamefont
  {Nayak}}]{Balents1998}%
  \BibitemOpen
  \bibfield  {author} {\bibinfo {author} {\bibfnamefont {L.}~\bibnamefont
  {Balents}}, \bibinfo {author} {\bibfnamefont {M.~P.}\ \bibnamefont {Fisher}},
  \ and\ \bibinfo {author} {\bibfnamefont {C.}~\bibnamefont {Nayak}},\
  }\href@noop {} {\bibfield  {journal} {\bibinfo  {journal} {Int. J. Mod. Phys.
  B}\ }\textbf {\bibinfo {volume} {12}},\ \bibinfo {pages} {1033} (\bibinfo
  {year} {1998})}\BibitemShut {NoStop}%
\bibitem [{\citenamefont {Read}\ and\ \citenamefont {Green}(2000)}]{Read2000}%
  \BibitemOpen
  \bibfield  {author} {\bibinfo {author} {\bibfnamefont {N.}~\bibnamefont
  {Read}}\ and\ \bibinfo {author} {\bibfnamefont {D.}~\bibnamefont {Green}},\
  }\href {\doibase 10.1103/PhysRevB.61.10267} {\bibfield  {journal} {\bibinfo
  {journal} {Phys. Rev. B}\ }\textbf {\bibinfo {volume} {61}},\ \bibinfo
  {pages} {10267} (\bibinfo {year} {2000})}\BibitemShut {NoStop}%
\bibitem [{\citenamefont {Hansson}\ \emph {et~al.}(2004)\citenamefont
  {Hansson}, \citenamefont {Oganesyan},\ and\ \citenamefont
  {Sondhi}}]{Hansson2004}%
  \BibitemOpen
  \bibfield  {author} {\bibinfo {author} {\bibfnamefont {T.~T.~H.}\
  \bibnamefont {Hansson}}, \bibinfo {author} {\bibfnamefont {V.}~\bibnamefont
  {Oganesyan}}, \ and\ \bibinfo {author} {\bibfnamefont {S.~S.~L.}\
  \bibnamefont {Sondhi}},\ }\href {\doibase 10.1016/j.aop.2004.05.006}
  {\bibfield  {journal} {\bibinfo  {journal} {Ann. Phys. (N. Y).}\ }\textbf
  {\bibinfo {volume} {313}},\ \bibinfo {pages} {497} (\bibinfo {year}
  {2004})}\BibitemShut {NoStop}%
\bibitem [{\citenamefont {Hansson}\ \emph {et~al.}(2012)\citenamefont
  {Hansson}, \citenamefont {Karlhede},\ and\ \citenamefont
  {Sato}}]{Hansson2012}%
  \BibitemOpen
  \bibfield  {author} {\bibinfo {author} {\bibfnamefont {T.~H.}\ \bibnamefont
  {Hansson}}, \bibinfo {author} {\bibfnamefont {A.}~\bibnamefont {Karlhede}}, \
  and\ \bibinfo {author} {\bibfnamefont {M.}~\bibnamefont {Sato}},\ }\href
  {\doibase 10.1088/1367-2630/14/6/063017} {\bibfield  {journal} {\bibinfo
  {journal} {New J. Phys.}\ }\textbf {\bibinfo {volume} {14}},\ \bibinfo
  {pages} {63017} (\bibinfo {year} {2012})},\ \Eprint
  {http://arxiv.org/abs/1105.5031} {arXiv:1105.5031 [cond-mat.supr-con]}
  \BibitemShut {NoStop}%
\bibitem [{\citenamefont {Hansson}\ \emph {et~al.}(2015)\citenamefont
  {Hansson}, \citenamefont {Kvorning}, \citenamefont {Nair},\ and\
  \citenamefont {Sreejith}}]{Hansson2015}%
  \BibitemOpen
  \bibfield  {author} {\bibinfo {author} {\bibfnamefont {T.~H.}\ \bibnamefont
  {Hansson}}, \bibinfo {author} {\bibfnamefont {T.}~\bibnamefont {Kvorning}},
  \bibinfo {author} {\bibfnamefont {V.~P.}\ \bibnamefont {Nair}}, \ and\
  \bibinfo {author} {\bibfnamefont {G.~J.}\ \bibnamefont {Sreejith}},\ }\href
  {\doibase 10.1103/PhysRevB.91.075116} {\bibfield  {journal} {\bibinfo
  {journal} {Phys. Rev. B}\ }\textbf {\bibinfo {volume} {91}},\ \bibinfo
  {pages} {075116} (\bibinfo {year} {2015})}\BibitemShut {NoStop}%
\bibitem [{\citenamefont {Vollhardt}\ and\ \citenamefont
  {Wolfle}(1990)}]{vollhardt}%
  \BibitemOpen
  \bibfield  {author} {\bibinfo {author} {\bibfnamefont {D.}~\bibnamefont
  {Vollhardt}}\ and\ \bibinfo {author} {\bibfnamefont {P.}~\bibnamefont
  {Wolfle}},\ }\href@noop {} {\emph {\bibinfo {title} {{Superfluid phases of
  helium 3}}}}\ (\bibinfo  {publisher} {Taylor and Francis Ltd},\ \bibinfo
  {year} {1990})\BibitemShut {NoStop}%
\bibitem [{\citenamefont {Volovik}(2009)}]{volovikbook}%
  \BibitemOpen
  \bibfield  {author} {\bibinfo {author} {\bibfnamefont {G.~E.}\ \bibnamefont
  {Volovik}},\ }\href@noop {} {\emph {\bibinfo {title} {{The universe in a
  helium droplet}}}},\ Vol.\ \bibinfo {volume} {117}\ (\bibinfo  {publisher}
  {Oxford University Press New York},\ \bibinfo {year} {2009})\BibitemShut
  {NoStop}%
\bibitem [{\citenamefont {{Kallin}}\ and\ \citenamefont
  {{Berlinsky}}()}]{Kallin2015}%
  \BibitemOpen
  \bibfield  {author} {\bibinfo {author} {\bibfnamefont {C.}~\bibnamefont
  {{Kallin}}}\ and\ \bibinfo {author} {\bibfnamefont {J.}~\bibnamefont
  {{Berlinsky}}},\ }\href@noop {} {\ }\Eprint {http://arxiv.org/abs/1512.01151}
  {arXiv:1512.01151 [cond-mat.supr-con]} \BibitemShut {NoStop}%
\bibitem [{\citenamefont {Gurarie}\ \emph {et~al.}(2005)\citenamefont
  {Gurarie}, \citenamefont {Radzihovsky},\ and\ \citenamefont
  {Andreev}}]{Gurarie2005}%
  \BibitemOpen
  \bibfield  {author} {\bibinfo {author} {\bibfnamefont {V.}~\bibnamefont
  {Gurarie}}, \bibinfo {author} {\bibfnamefont {L.}~\bibnamefont
  {Radzihovsky}}, \ and\ \bibinfo {author} {\bibfnamefont {A.~V.}\ \bibnamefont
  {Andreev}},\ }\href {\doibase 10.1103/PhysRevLett.94.230403} {\bibfield
  {journal} {\bibinfo  {journal} {Phys. Rev. Lett.}\ }\textbf {\bibinfo
  {volume} {94}},\ \bibinfo {pages} {230403} (\bibinfo {year}
  {2005})}\BibitemShut {NoStop}%
\bibitem [{\citenamefont {Gurarie}\ and\ \citenamefont
  {Radzihovsky}(2007)}]{Gurarie20072}%
  \BibitemOpen
  \bibfield  {author} {\bibinfo {author} {\bibfnamefont {V.}~\bibnamefont
  {Gurarie}}\ and\ \bibinfo {author} {\bibfnamefont {L.}~\bibnamefont
  {Radzihovsky}},\ }\href {\doibase
  http://dx.doi.org/10.1016/j.aop.2006.10.009} {\bibfield  {journal} {\bibinfo
  {journal} {Ann. Phys. (N. Y).}\ }\textbf {\bibinfo {volume} {322}},\ \bibinfo
  {pages} {2 } (\bibinfo {year} {2007})}\BibitemShut {NoStop}%
\bibitem [{\citenamefont {Nayak}\ \emph {et~al.}(2008)\citenamefont {Nayak},
  \citenamefont {Simon}, \citenamefont {Stern}, \citenamefont {Freedman},\ and\
  \citenamefont {Sarma}}]{nayak2008}%
  \BibitemOpen
  \bibfield  {author} {\bibinfo {author} {\bibfnamefont {C.}~\bibnamefont
  {Nayak}}, \bibinfo {author} {\bibfnamefont {S.~H.}\ \bibnamefont {Simon}},
  \bibinfo {author} {\bibfnamefont {A.}~\bibnamefont {Stern}}, \bibinfo
  {author} {\bibfnamefont {M.}~\bibnamefont {Freedman}}, \ and\ \bibinfo
  {author} {\bibfnamefont {S.~D.}\ \bibnamefont {Sarma}},\ }\href@noop {}
  {\bibfield  {journal} {\bibinfo  {journal} {Rev. Mod. Phys.}\ }\textbf
  {\bibinfo {volume} {80}},\ \bibinfo {pages} {1083} (\bibinfo {year}
  {2008})}\BibitemShut {NoStop}%
\bibitem [{\citenamefont {Ryu}\ \emph {et~al.}(2010)\citenamefont {Ryu},
  \citenamefont {Schnyder}, \citenamefont {Furusaki},\ and\ \citenamefont
  {Ludwig}}]{ryu2010}%
  \BibitemOpen
  \bibfield  {author} {\bibinfo {author} {\bibfnamefont {S.}~\bibnamefont
  {Ryu}}, \bibinfo {author} {\bibfnamefont {A.~P.}\ \bibnamefont {Schnyder}},
  \bibinfo {author} {\bibfnamefont {A.}~\bibnamefont {Furusaki}}, \ and\
  \bibinfo {author} {\bibfnamefont {A.~W.~W.}\ \bibnamefont {Ludwig}},\ }\href
  {http://stacks.iop.org/1367-2630/12/i=6/a=065010} {\bibfield  {journal}
  {\bibinfo  {journal} {New J. Phys.}\ }\textbf {\bibinfo {volume} {12}},\
  \bibinfo {pages} {065010} (\bibinfo {year} {2010})}\BibitemShut {NoStop}%
\bibitem [{\citenamefont {Senthil}\ \emph {et~al.}(1999)\citenamefont
  {Senthil}, \citenamefont {Marston},\ and\ \citenamefont
  {Fisher}}]{Senthil1999}%
  \BibitemOpen
  \bibfield  {author} {\bibinfo {author} {\bibfnamefont {T.}~\bibnamefont
  {Senthil}}, \bibinfo {author} {\bibfnamefont {J.}~\bibnamefont {Marston}}, \
  and\ \bibinfo {author} {\bibfnamefont {M.}~\bibnamefont {Fisher}},\ }\href
  {\doibase 10.1103/PhysRevB.60.4245} {\bibfield  {journal} {\bibinfo
  {journal} {Phys. Rev. B}\ }\textbf {\bibinfo {volume} {60}},\ \bibinfo
  {pages} {4245} (\bibinfo {year} {1999})}\BibitemShut {NoStop}%
\bibitem [{\citenamefont {Teo}\ and\ \citenamefont {Kane}(2010)}]{Teo2010}%
  \BibitemOpen
  \bibfield  {author} {\bibinfo {author} {\bibfnamefont {J.~C.~Y.}\
  \bibnamefont {Teo}}\ and\ \bibinfo {author} {\bibfnamefont {C.~L.}\
  \bibnamefont {Kane}},\ }\href {\doibase 10.1103/PhysRevB.82.115120}
  {\bibfield  {journal} {\bibinfo  {journal} {Phys. Rev. B}\ }\textbf {\bibinfo
  {volume} {82}},\ \bibinfo {pages} {1} (\bibinfo {year} {2010})}\BibitemShut
  {NoStop}%
\bibitem [{\citenamefont {Tada}\ \emph {et~al.}(2015)\citenamefont {Tada},
  \citenamefont {Nie},\ and\ \citenamefont {Oshikawa}}]{Tada2015}%
  \BibitemOpen
  \bibfield  {author} {\bibinfo {author} {\bibfnamefont {Y.}~\bibnamefont
  {Tada}}, \bibinfo {author} {\bibfnamefont {W.}~\bibnamefont {Nie}}, \ and\
  \bibinfo {author} {\bibfnamefont {M.}~\bibnamefont {Oshikawa}},\ }\href
  {\doibase 10.1103/PhysRevLett.114.195301} {\bibfield  {journal} {\bibinfo
  {journal} {Phys. Rev. Lett.}\ }\textbf {\bibinfo {volume} {114}},\ \bibinfo
  {pages} {195301} (\bibinfo {year} {2015})}\BibitemShut {NoStop}%
\bibitem [{\citenamefont {Volovik}(2014)}]{Volovik2014a}%
  \BibitemOpen
  \bibfield  {author} {\bibinfo {author} {\bibfnamefont {G.~E.}\ \bibnamefont
  {Volovik}},\ }\href {\doibase 10.7868/S0370274X14230118} {\bibfield
  {journal} {\bibinfo  {journal} {JETP Lett.}\ }\textbf {\bibinfo {volume}
  {100}},\ \bibinfo {pages} {742} (\bibinfo {year} {2014})}\BibitemShut
  {NoStop}%
\bibitem [{\citenamefont {Nishikubo}\ \emph {et~al.}(2011)\citenamefont
  {Nishikubo}, \citenamefont {Kudo},\ and\ \citenamefont
  {Nohara}}]{Nishikubo2011}%
  \BibitemOpen
  \bibfield  {author} {\bibinfo {author} {\bibfnamefont {Y.}~\bibnamefont
  {Nishikubo}}, \bibinfo {author} {\bibfnamefont {K.}~\bibnamefont {Kudo}}, \
  and\ \bibinfo {author} {\bibfnamefont {M.}~\bibnamefont {Nohara}},\ }\href
  {\doibase 10.1143/JPSJ.80.055002} {\bibfield  {journal} {\bibinfo  {journal}
  {J. Phys. Soc. Japan}\ }\textbf {\bibinfo {volume} {80}},\ \bibinfo {pages}
  {055002} (\bibinfo {year} {2011})}\BibitemShut {NoStop}%
\bibitem [{\citenamefont {Fischer}\ \emph {et~al.}(2014)\citenamefont
  {Fischer}, \citenamefont {Neupert}, \citenamefont {Platt}, \citenamefont
  {Schnyder}, \citenamefont {Hanke}, \citenamefont {Goryo}, \citenamefont
  {Thomale},\ and\ \citenamefont {Sigrist}}]{Fischer2014}%
  \BibitemOpen
  \bibfield  {author} {\bibinfo {author} {\bibfnamefont {M.~H.}\ \bibnamefont
  {Fischer}}, \bibinfo {author} {\bibfnamefont {T.}~\bibnamefont {Neupert}},
  \bibinfo {author} {\bibfnamefont {C.}~\bibnamefont {Platt}}, \bibinfo
  {author} {\bibfnamefont {A.~P.}\ \bibnamefont {Schnyder}}, \bibinfo {author}
  {\bibfnamefont {W.}~\bibnamefont {Hanke}}, \bibinfo {author} {\bibfnamefont
  {J.}~\bibnamefont {Goryo}}, \bibinfo {author} {\bibfnamefont
  {R.}~\bibnamefont {Thomale}}, \ and\ \bibinfo {author} {\bibfnamefont
  {M.}~\bibnamefont {Sigrist}},\ }\href {\doibase 10.1103/PhysRevB.89.020509}
  {\bibfield  {journal} {\bibinfo  {journal} {Phys. Rev. B}\ }\textbf {\bibinfo
  {volume} {89}},\ \bibinfo {pages} {020509} (\bibinfo {year}
  {2014})}\BibitemShut {NoStop}%
\bibitem [{\citenamefont {Nandkishore}\ \emph {et~al.}(2012)\citenamefont
  {Nandkishore}, \citenamefont {Levitov},\ and\ \citenamefont
  {Chubukov}}]{nandkishore2012}%
  \BibitemOpen
  \bibfield  {author} {\bibinfo {author} {\bibfnamefont {R.}~\bibnamefont
  {Nandkishore}}, \bibinfo {author} {\bibfnamefont {L.}~\bibnamefont
  {Levitov}}, \ and\ \bibinfo {author} {\bibfnamefont {A.}~\bibnamefont
  {Chubukov}},\ }\href@noop {} {\bibfield  {journal} {\bibinfo  {journal}
  {Nature Physics}\ }\textbf {\bibinfo {volume} {8}},\ \bibinfo {pages} {158}
  (\bibinfo {year} {2012})}\BibitemShut {NoStop}%
\bibitem [{\citenamefont {Black-Schaffer}\ and\ \citenamefont
  {Honerkamp}(2014)}]{Honerkamp2014}%
  \BibitemOpen
  \bibfield  {author} {\bibinfo {author} {\bibfnamefont {A.~M.}\ \bibnamefont
  {Black-Schaffer}}\ and\ \bibinfo {author} {\bibfnamefont {C.}~\bibnamefont
  {Honerkamp}},\ }\href {http://stacks.iop.org/0953-8984/26/i=42/a=423201}
  {\bibfield  {journal} {\bibinfo  {journal} {J. Phys.}\ }\textbf {\bibinfo
  {volume} {26}},\ \bibinfo {pages} {423201} (\bibinfo {year}
  {2014})}\BibitemShut {NoStop}%
\bibitem [{\citenamefont {Kiesel}\ \emph {et~al.}(2012)\citenamefont {Kiesel},
  \citenamefont {Platt}, \citenamefont {Hanke}, \citenamefont {Abanin},\ and\
  \citenamefont {Thomale}}]{Kiesel2012}%
  \BibitemOpen
  \bibfield  {author} {\bibinfo {author} {\bibfnamefont {M.~L.}\ \bibnamefont
  {Kiesel}}, \bibinfo {author} {\bibfnamefont {C.}~\bibnamefont {Platt}},
  \bibinfo {author} {\bibfnamefont {W.}~\bibnamefont {Hanke}}, \bibinfo
  {author} {\bibfnamefont {D.~A.}\ \bibnamefont {Abanin}}, \ and\ \bibinfo
  {author} {\bibfnamefont {R.}~\bibnamefont {Thomale}},\ }\href {\doibase
  10.1103/PhysRevB.86.020507} {\bibfield  {journal} {\bibinfo  {journal} {Phys.
  Rev. B}\ }\textbf {\bibinfo {volume} {86}},\ \bibinfo {pages} {020507}
  (\bibinfo {year} {2012})}\BibitemShut {NoStop}%
\bibitem [{\citenamefont {Chen}\ \emph {et~al.}(2013)\citenamefont {Chen},
  \citenamefont {Gu}, \citenamefont {Liu},\ and\ \citenamefont
  {Wen}}]{Chen2013}%
  \BibitemOpen
  \bibfield  {author} {\bibinfo {author} {\bibfnamefont {X.}~\bibnamefont
  {Chen}}, \bibinfo {author} {\bibfnamefont {Z.~C.}\ \bibnamefont {Gu}},
  \bibinfo {author} {\bibfnamefont {Z.~X.}\ \bibnamefont {Liu}}, \ and\
  \bibinfo {author} {\bibfnamefont {X.~G.}\ \bibnamefont {Wen}},\ }\href
  {\doibase 10.1103/PhysRevB.87.155114} {\bibfield  {journal} {\bibinfo
  {journal} {Phys. Rev. B}\ }\textbf {\bibinfo {volume} {87}},\ \bibinfo
  {pages} {1} (\bibinfo {year} {2013})}\BibitemShut {NoStop}%
\bibitem [{\citenamefont {{Keldysh}}(1979)}]{Keldysh1979}%
  \BibitemOpen
  \bibfield  {author} {\bibinfo {author} {\bibfnamefont {L.~V.}\ \bibnamefont
  {{Keldysh}}},\ }\href@noop {} {\bibfield  {journal} {\bibinfo  {journal}
  {JETP Lett.}\ }\textbf {\bibinfo {volume} {29}},\ \bibinfo {pages} {658}
  (\bibinfo {year} {1979})}\BibitemShut {NoStop}%
\bibitem [{\citenamefont {Kitaev}(2006)}]{Kitaev2006}%
  \BibitemOpen
  \bibfield  {author} {\bibinfo {author} {\bibfnamefont {A.}~\bibnamefont
  {Kitaev}},\ }\href {\doibase 10.1016/j.aop.2005.10.005} {\bibfield  {journal}
  {\bibinfo  {journal} {Ann. Phys. (N. Y).}\ }\textbf {\bibinfo {volume}
  {321}},\ \bibinfo {pages} {2} (\bibinfo {year} {2006})}\BibitemShut {NoStop}%
\bibitem [{Note1()}]{Note1}%
  \BibitemOpen
  \bibinfo {note} {An \protect \emph {abelian phase} is characterized by a
  unique anyon resulting from the fusion of any two excitations. On the other
  hand, \protect \emph {non-abelian} anyons can fuse into different
  outcomes.}\BibitemShut {Stop}%
\bibitem [{\citenamefont {Read}(1990)}]{Read1990}%
  \BibitemOpen
  \bibfield  {author} {\bibinfo {author} {\bibfnamefont {N.}~\bibnamefont
  {Read}},\ }\href {\doibase 10.1103/PhysRevLett.65.1502} {\bibfield  {journal}
  {\bibinfo  {journal} {Phys. Rev. Lett.}\ }\textbf {\bibinfo {volume} {65}},\
  \bibinfo {pages} {1502} (\bibinfo {year} {1990})}\BibitemShut {NoStop}%
\bibitem [{\citenamefont {Wen}\ and\ \citenamefont
  {Zee}(1992{\natexlab{a}})}]{Wen1992b}%
  \BibitemOpen
  \bibfield  {author} {\bibinfo {author} {\bibfnamefont {X.-G.}\ \bibnamefont
  {Wen}}\ and\ \bibinfo {author} {\bibfnamefont {A.}~\bibnamefont {Zee}},\
  }\href {http://journals.aps.org/prb/abstract/10.1103/PhysRevB.46.2290}
  {\bibfield  {journal} {\bibinfo  {journal} {Phys. Rev. B}\ }\textbf {\bibinfo
  {volume} {46}},\ \bibinfo {pages} {2290} (\bibinfo {year}
  {1992}{\natexlab{a}})}\BibitemShut {NoStop}%
\bibitem [{\citenamefont {Frohlich}\ and\ \citenamefont
  {Zee}(1991)}]{Frohlich1991}%
  \BibitemOpen
  \bibfield  {author} {\bibinfo {author} {\bibfnamefont {J.}~\bibnamefont
  {Frohlich}}\ and\ \bibinfo {author} {\bibfnamefont {A.}~\bibnamefont {Zee}},\
  }\href {\doibase http://dx.doi.org/10.1016/0550-3213(91)90275-3} {\bibfield
  {journal} {\bibinfo  {journal} {Nucl. Phys. B}\ }\textbf {\bibinfo {volume}
  {364}},\ \bibinfo {pages} {517 } (\bibinfo {year} {1991})}\BibitemShut
  {NoStop}%
\bibitem [{Note2()}]{Note2}%
  \BibitemOpen
  \bibinfo {note} {A remark regarding \protect \emph {compactness}: The
  statistical fields dual to global conserved currents are non-compact which
  ensures absence of instanton magnetic monopoles and strict conservation of
  these currents. On the other hand, compact statistical gauge fields are also
  present sometimes, but these do not encode any strict conservation
  laws.}\BibitemShut {Stop}%
\bibitem [{\citenamefont {Wen}(1995)}]{Wen1995}%
  \BibitemOpen
  \bibfield  {author} {\bibinfo {author} {\bibfnamefont {X.-G.}\ \bibnamefont
  {Wen}},\ }\href {\doibase 10.1080/00018739500101566} {\bibfield  {journal}
  {\bibinfo  {journal} {Adv. Phys.}\ }\textbf {\bibinfo {volume} {44}},\
  \bibinfo {pages} {405} (\bibinfo {year} {1995})}\BibitemShut {NoStop}%
\bibitem [{\citenamefont {Lu}\ and\ \citenamefont {Vishwanath}(2012)}]{Lu2012}%
  \BibitemOpen
  \bibfield  {author} {\bibinfo {author} {\bibfnamefont {Y.~M.}\ \bibnamefont
  {Lu}}\ and\ \bibinfo {author} {\bibfnamefont {A.}~\bibnamefont
  {Vishwanath}},\ }\href {\doibase 10.1103/PhysRevB.86.125119} {\bibfield
  {journal} {\bibinfo  {journal} {Phys. Rev. B}\ }\textbf {\bibinfo {volume}
  {86}},\ \bibinfo {pages} {1} (\bibinfo {year} {2012})}\BibitemShut {NoStop}%
\bibitem [{Note3()}]{Note3}%
  \BibitemOpen
  \bibinfo {note} {Indeed, the number of entries of the $K$-matrix might be
  much larger than the number of independent braiding phases.}\BibitemShut
  {Stop}%
\bibitem [{\citenamefont {Wen}(1992)}]{Wen1992a}%
  \BibitemOpen
  \bibfield  {author} {\bibinfo {author} {\bibfnamefont {X.-G.}\ \bibnamefont
  {Wen}},\ }\href {\doibase 10.1142/S0217979292000840} {\bibfield  {journal}
  {\bibinfo  {journal} {Int. J. Mod. Phys. B}\ }\textbf {\bibinfo {volume}
  {06}},\ \bibinfo {pages} {1711} (\bibinfo {year} {1992})}\BibitemShut
  {NoStop}%
\bibitem [{\citenamefont {Haldane}(1995)}]{Haldane1995}%
  \BibitemOpen
  \bibfield  {author} {\bibinfo {author} {\bibfnamefont {F.~D.~M.}\
  \bibnamefont {Haldane}},\ }\href {\doibase 10.1103/PhysRevLett.74.2090}
  {\bibfield  {journal} {\bibinfo  {journal} {Phys. Rev. Lett.}\ }\textbf
  {\bibinfo {volume} {74}},\ \bibinfo {pages} {2090} (\bibinfo {year}
  {1995})}\BibitemShut {NoStop}%
\bibitem [{\citenamefont {Levin}(2013)}]{Levin2013}%
  \BibitemOpen
  \bibfield  {author} {\bibinfo {author} {\bibfnamefont {M.}~\bibnamefont
  {Levin}},\ }\href {\doibase 10.1103/PhysRevX.3.021009} {\bibfield  {journal}
  {\bibinfo  {journal} {Phys. Rev. X}\ }\textbf {\bibinfo {volume} {3}},\
  \bibinfo {pages} {021009} (\bibinfo {year} {2013})}\BibitemShut {NoStop}%
\bibitem [{\citenamefont {Kovner}\ \emph {et~al.}(1991)\citenamefont {Kovner},
  \citenamefont {Rosenstein},\ and\ \citenamefont {Eliezer}}]{Kovner1991}%
  \BibitemOpen
  \bibfield  {author} {\bibinfo {author} {\bibfnamefont {A.}~\bibnamefont
  {Kovner}}, \bibinfo {author} {\bibfnamefont {B.}~\bibnamefont {Rosenstein}},
  \ and\ \bibinfo {author} {\bibfnamefont {D.}~\bibnamefont {Eliezer}},\ }\href
  {http://www.sciencedirect.com/science/article/pii/055032139190263W}
  {\bibfield  {journal} {\bibinfo  {journal} {Nucl. Phys. B}\ }\textbf
  {\bibinfo {volume} {350}},\ \bibinfo {pages} {325} (\bibinfo {year}
  {1991})}\BibitemShut {NoStop}%
\bibitem [{Note4()}]{Note4}%
  \BibitemOpen
  \bibinfo {note} {\label {fa} For simplicity we set the mass of an elementary
  fermion to unity. Its electric charge $e$ is set to \protect \emph {minus}
  unity, which fixes the magnetic flux carried by an elementary
  (counterclockwise) vortex $\varphi (\protect \mathbf {x})=\protect \text
  {arg}(\protect \mathbf {x})/2$ to $+\pi $, i.e., a half of a magnetic flux
  quantum.}\BibitemShut {Stop}%
\bibitem [{Note5()}]{Note5}%
  \BibitemOpen
  \bibinfo {note} {Here the indices are raised and lowered with the Minkowski
  metric.}\BibitemShut {Stop}%
\bibitem [{\citenamefont {Balents}\ \emph {et~al.}(1999)\citenamefont
  {Balents}, \citenamefont {Fisher},\ and\ \citenamefont
  {Nayak}}]{Balents1999}%
  \BibitemOpen
  \bibfield  {author} {\bibinfo {author} {\bibfnamefont {L.}~\bibnamefont
  {Balents}}, \bibinfo {author} {\bibfnamefont {M.~P.~A.}\ \bibnamefont
  {Fisher}}, \ and\ \bibinfo {author} {\bibfnamefont {C.}~\bibnamefont
  {Nayak}},\ }\href {\doibase 10.1103/PhysRevB.60.1654} {\bibfield  {journal}
  {\bibinfo  {journal} {Phys. Rev. B}\ }\textbf {\bibinfo {volume} {60}},\
  \bibinfo {pages} {1654} (\bibinfo {year} {1999})}\BibitemShut {NoStop}%
\bibitem [{\citenamefont {{Hermanns}}()}]{Hermanns2008}%
  \BibitemOpen
  \bibfield  {author} {\bibinfo {author} {\bibfnamefont {M.}~\bibnamefont
  {{Hermanns}}},\ }\href@noop {} {\ }\Eprint {http://arxiv.org/abs/0804.1332}
  {arXiv:0804.1332 [cond-mat.supr-con]} \BibitemShut {NoStop}%
\bibitem [{Note6()}]{Note6}%
  \BibitemOpen
  \bibinfo {note} {Indeed, across the vortex branch cut, the phase of every
  Dirac component of the spinor $\protect \mathaccentV {tilde}07E\psi _i=
  \protect \qopname \relax o{exp}(-i \varphi \tau _z) \psi _i$ changes by $\pm
  \pi $.}\BibitemShut {Stop}%
\bibitem [{\citenamefont {{Anderson}}(1998)}]{Anderson1998}%
  \BibitemOpen
  \bibfield  {author} {\bibinfo {author} {\bibfnamefont {P.~W.}\ \bibnamefont
  {{Anderson}}},\ }\href@noop {} {\bibfield  {journal} {\bibinfo  {journal}
  {arXiv:cond-mat/9812063}\ } (\bibinfo {year} {1998})}\BibitemShut {NoStop}%
\bibitem [{\citenamefont {Ariad}\ \emph {et~al.}(2015)\citenamefont {Ariad},
  \citenamefont {Grosfeld},\ and\ \citenamefont {Seradjeh}}]{Ariad2015}%
  \BibitemOpen
  \bibfield  {author} {\bibinfo {author} {\bibfnamefont {D.}~\bibnamefont
  {Ariad}}, \bibinfo {author} {\bibfnamefont {E.}~\bibnamefont {Grosfeld}}, \
  and\ \bibinfo {author} {\bibfnamefont {B.}~\bibnamefont {Seradjeh}},\ }\href
  {\doibase 10.1103/PhysRevB.92.035136} {\bibfield  {journal} {\bibinfo
  {journal} {Phys. Rev. B}\ }\textbf {\bibinfo {volume} {92}},\ \bibinfo
  {pages} {035136} (\bibinfo {year} {2015})}\BibitemShut {NoStop}%
\bibitem [{Note7()}]{Note7}%
  \BibitemOpen
  \bibinfo {note} {To this end, perform the following transformation of the
  last term in Eq. \protect \textup {\hbox {\mathsurround \z@ \protect
  \normalfont (\ignorespaces \ref {dxy1}\unskip \@@italiccorr )}}: first rotate
  by a $\pi /2$ angle, i.e., $X\to -Y$, $Y\to +X$ and second apply a unitary
  rotation $U= \protect \qopname \relax o{exp}(i \pi \tau _z)$ to $\protect
  \mathaccentV {tilde}07E\psi _2$.}\BibitemShut {Stop}%
\bibitem [{Note8()}]{Note8}%
  \BibitemOpen
  \bibinfo {note} {It would be instructive to derive Eqs. \protect \textup
  {\hbox {\mathsurround \z@ \protect \normalfont (\ignorespaces \ref
  {topfers}\unskip \@@italiccorr )}}, \protect \textup {\hbox {\mathsurround
  \z@ \protect \normalfont (\ignorespaces \ref {topferdid}\unskip \@@italiccorr
  )}} rigorously by using the functional bosonization approach developed in
  \cite {Chan2013}.}\BibitemShut {Stop}%
\bibitem [{Note9()}]{Note9}%
  \BibitemOpen
  \bibinfo {note} {By definition, an elementary excitation has trivial mutual
  statistics with all anyons.}\BibitemShut {Stop}%
\bibitem [{Note10()}]{Note10}%
  \BibitemOpen
  \bibinfo {note} {Specifically, in Fig. \ref {fig1a} we chose to identify
  $\epsilon $ with the spin-up Bogoliubov quasiparticle that carries the
  current $j_{\delimiter "3222378 }$ in Eq. \protect \textup {\hbox
  {\mathsurround \z@ \protect \normalfont (\ignorespaces \ref {topfers}\unskip
  \@@italiccorr )}}. Alternatively, in the $s$-wave case one can choose
  $l_{\epsilon \delimiter "3223379 }=(0,0,0, -1)$ which corresponds to the
  spin-down Bogoliubov quasiparticle. The latter identification gives the same
  braiding and fusion rules as the former, but obviously differs by the sign of
  the spin charge. Note that for the $d+id$ state, Eq. \protect \textup {\hbox
  {\mathsurround \z@ \protect \normalfont (\ignorespaces \ref
  {topferdid}\unskip \@@italiccorr )}} leads to $l_{\epsilon \delimiter
  "3223379 }=(0,0,0, 1)$. \label {footupdown}}\BibitemShut {NoStop}%
\bibitem [{\citenamefont {Reznik}\ and\ \citenamefont
  {Aharonov}(1989)}]{Reznik1989}%
  \BibitemOpen
  \bibfield  {author} {\bibinfo {author} {\bibfnamefont {B.}~\bibnamefont
  {Reznik}}\ and\ \bibinfo {author} {\bibfnamefont {Y.}~\bibnamefont
  {Aharonov}},\ }\href@noop {} {\bibfield  {journal} {\bibinfo  {journal}
  {Phys. Rev. D}\ }\textbf {\bibinfo {volume} {40}},\ \bibinfo {pages} {2112}
  (\bibinfo {year} {1989})}\BibitemShut {NoStop}%
\bibitem [{\citenamefont {Goldhaber}\ and\ \citenamefont
  {Kivelson}(1991)}]{Goldhaber1991}%
  \BibitemOpen
  \bibfield  {author} {\bibinfo {author} {\bibfnamefont {A.~S.}\ \bibnamefont
  {Goldhaber}}\ and\ \bibinfo {author} {\bibfnamefont {S.}~\bibnamefont
  {Kivelson}},\ }\href {\doibase
  http://dx.doi.org/10.1016/0370-2693(91)90792-O} {\bibfield  {journal}
  {\bibinfo  {journal} {Phys. Lett. B}\ }\textbf {\bibinfo {volume} {255}},\
  \bibinfo {pages} {445 } (\bibinfo {year} {1991})}\BibitemShut {NoStop}%
\bibitem [{\citenamefont {Bravyi}\ and\ \citenamefont {Kitaev}()}]{Bravyi1998}%
  \BibitemOpen
  \bibfield  {author} {\bibinfo {author} {\bibfnamefont {S.~B.}\ \bibnamefont
  {Bravyi}}\ and\ \bibinfo {author} {\bibfnamefont {A.~Y.}\ \bibnamefont
  {Kitaev}},\ }\href {http://xxx.lanl.gov/abs/quant-ph/9811052} {\ }\Eprint
  {http://arxiv.org/abs/9811052} {arXiv:9811052 [quant-ph]} \BibitemShut
  {NoStop}%
\bibitem [{Note11()}]{Note11}%
  \BibitemOpen
  \bibinfo {note} {Restoring $\hbar $, in our convention the unit of the spin
  charge is equal to $\hbar /2$ and thus $\sigma _s=\nu _{s} (\hbar /2)^2/h=
  \hbar /(4\pi )$}\BibitemShut {NoStop}%
\bibitem [{\citenamefont {{Bernevig}}\ and\ \citenamefont
  {{Neupert}}()}]{Bernevig2015}%
  \BibitemOpen
  \bibfield  {author} {\bibinfo {author} {\bibfnamefont {A.}~\bibnamefont
  {{Bernevig}}}\ and\ \bibinfo {author} {\bibfnamefont {T.}~\bibnamefont
  {{Neupert}}},\ }\href@noop {} {\ }\Eprint {http://arxiv.org/abs/1506.05805}
  {arXiv:1506.05805 [cond-mat.str-el]} \BibitemShut {NoStop}%
\bibitem [{\citenamefont {Kane}\ and\ \citenamefont {Fisher}(1997)}]{Kane1997}%
  \BibitemOpen
  \bibfield  {author} {\bibinfo {author} {\bibfnamefont {C.~L.}\ \bibnamefont
  {Kane}}\ and\ \bibinfo {author} {\bibfnamefont {M.~P.~A.}\ \bibnamefont
  {Fisher}},\ }\href {\doibase 10.1103/PhysRevB.55.15832} {\bibfield  {journal}
  {\bibinfo  {journal} {Phys. Rev. B}\ }\textbf {\bibinfo {volume} {55}},\
  \bibinfo {pages} {15832} (\bibinfo {year} {1997})}\BibitemShut {NoStop}%
\bibitem [{\citenamefont {Ginsparg}(1988)}]{Ginsparg1988}%
  \BibitemOpen
  \bibfield  {author} {\bibinfo {author} {\bibfnamefont {P.}~\bibnamefont
  {Ginsparg}},\ }\href {http://arxiv.org/abs/hep-th/9108028} {\bibfield
  {journal} {\bibinfo  {journal} {Physics (College. Park. Md).}\ }\textbf
  {\bibinfo {volume} {88}},\ \bibinfo {pages} {90} (\bibinfo {year} {1988})},\
  \Eprint {http://arxiv.org/abs/9108028v1} {arXiv:9108028v1 [arXiv:hep-th]}
  \BibitemShut {NoStop}%
\bibitem [{\citenamefont {You}\ \emph {et~al.}(2015)\citenamefont {You},
  \citenamefont {Bi}, \citenamefont {Rasmussen}, \citenamefont {Cheng},\ and\
  \citenamefont {Xu}}]{You2015}%
  \BibitemOpen
  \bibfield  {author} {\bibinfo {author} {\bibfnamefont {Y.-Z.}\ \bibnamefont
  {You}}, \bibinfo {author} {\bibfnamefont {Z.}~\bibnamefont {Bi}}, \bibinfo
  {author} {\bibfnamefont {A.}~\bibnamefont {Rasmussen}}, \bibinfo {author}
  {\bibfnamefont {M.}~\bibnamefont {Cheng}}, \ and\ \bibinfo {author}
  {\bibfnamefont {C.}~\bibnamefont {Xu}},\ }\href {\doibase
  10.1088/1367-2630/17/7/075010} {\bibfield  {journal} {\bibinfo  {journal}
  {New J. Phys.}\ }\textbf {\bibinfo {volume} {17}},\ \bibinfo {pages} {075010}
  (\bibinfo {year} {2015})}\BibitemShut {NoStop}%
\bibitem [{\citenamefont {Cano}\ \emph {et~al.}(2014)\citenamefont {Cano},
  \citenamefont {Cheng}, \citenamefont {Mulligan}, \citenamefont {Nayak},
  \citenamefont {Plamadeala},\ and\ \citenamefont {Yard}}]{Cano2014}%
  \BibitemOpen
  \bibfield  {author} {\bibinfo {author} {\bibfnamefont {J.}~\bibnamefont
  {Cano}}, \bibinfo {author} {\bibfnamefont {M.}~\bibnamefont {Cheng}},
  \bibinfo {author} {\bibfnamefont {M.}~\bibnamefont {Mulligan}}, \bibinfo
  {author} {\bibfnamefont {C.}~\bibnamefont {Nayak}}, \bibinfo {author}
  {\bibfnamefont {E.}~\bibnamefont {Plamadeala}}, \ and\ \bibinfo {author}
  {\bibfnamefont {J.}~\bibnamefont {Yard}},\ }\href {\doibase
  10.1103/PhysRevB.89.115116} {\bibfield  {journal} {\bibinfo  {journal} {Phys.
  Rev. B}\ }\textbf {\bibinfo {volume} {89}},\ \bibinfo {pages} {115116}
  (\bibinfo {year} {2014})}\BibitemShut {NoStop}%
\bibitem [{Note12()}]{Note12}%
  \BibitemOpen
  \bibinfo {note} {Incidentally, these purely bosonic Cartan $K$-matrices have
  also been proposed for describing $s$-wave superconductors strongly
  proximity-coupled to topological Chern insulators in \cite
  {Teo2015}.}\BibitemShut {Stop}%
\bibitem [{\citenamefont {Bonderson}\ and\ \citenamefont
  {Nayak}(2013)}]{Bonderson2013}%
  \BibitemOpen
  \bibfield  {author} {\bibinfo {author} {\bibfnamefont {P.}~\bibnamefont
  {Bonderson}}\ and\ \bibinfo {author} {\bibfnamefont {C.}~\bibnamefont
  {Nayak}},\ }\href {\doibase 10.1103/PhysRevB.87.195451} {\bibfield  {journal}
  {\bibinfo  {journal} {Phys. Rev. B}\ }\textbf {\bibinfo {volume} {87}},\
  \bibinfo {pages} {195451} (\bibinfo {year} {2013})}\BibitemShut {NoStop}%
\bibitem [{\citenamefont {Polyakov}(1987)}]{Polyakovbook}%
  \BibitemOpen
  \bibfield  {author} {\bibinfo {author} {\bibfnamefont {A.~M.}\ \bibnamefont
  {Polyakov}},\ }\href {https://books.google.com/books?id=uaI8xcjJ8LMC} {\emph
  {\bibinfo {title} {Gauge Fields and Strings}}},\ Contemporary concepts in
  physics\ (\bibinfo  {publisher} {Taylor \& Francis},\ \bibinfo {year}
  {1987})\BibitemShut {NoStop}%
\bibitem [{\citenamefont {Fradkin}\ and\ \citenamefont
  {Shenker}(1979)}]{Fradkin1979}%
  \BibitemOpen
  \bibfield  {author} {\bibinfo {author} {\bibfnamefont {E.}~\bibnamefont
  {Fradkin}}\ and\ \bibinfo {author} {\bibfnamefont {S.~H.}\ \bibnamefont
  {Shenker}},\ }\href {\doibase 10.1103/PhysRevD.19.3682} {\bibfield  {journal}
  {\bibinfo  {journal} {Phys. Rev. D}\ }\textbf {\bibinfo {volume} {19}},\
  \bibinfo {pages} {3682} (\bibinfo {year} {1979})}\BibitemShut {NoStop}%
\bibitem [{\citenamefont {Metlitski}\ \emph {et~al.}(2013)\citenamefont
  {Metlitski}, \citenamefont {Kane},\ and\ \citenamefont
  {Fisher}}]{Metlitski2013}%
  \BibitemOpen
  \bibfield  {author} {\bibinfo {author} {\bibfnamefont {M.~A.}\ \bibnamefont
  {Metlitski}}, \bibinfo {author} {\bibfnamefont {C.~L.}\ \bibnamefont {Kane}},
  \ and\ \bibinfo {author} {\bibfnamefont {M.~P.~A.}\ \bibnamefont {Fisher}},\
  }\href {\doibase 10.1103/PhysRevB.88.035131} {\bibfield  {journal} {\bibinfo
  {journal} {Phys. Rev. B}\ }\textbf {\bibinfo {volume} {88}},\ \bibinfo
  {pages} {035131} (\bibinfo {year} {2013})}\BibitemShut {NoStop}%
\bibitem [{\citenamefont {Wang}\ and\ \citenamefont
  {Senthil}(2013)}]{Chong2013}%
  \BibitemOpen
  \bibfield  {author} {\bibinfo {author} {\bibfnamefont {C.}~\bibnamefont
  {Wang}}\ and\ \bibinfo {author} {\bibfnamefont {T.}~\bibnamefont {Senthil}},\
  }\href {\doibase 10.1103/PhysRevB.87.235122} {\bibfield  {journal} {\bibinfo
  {journal} {Phys. Rev. B}\ }\textbf {\bibinfo {volume} {87}},\ \bibinfo
  {pages} {235122} (\bibinfo {year} {2013})}\BibitemShut {NoStop}%
\bibitem [{\citenamefont {Read}(2009)}]{Read2009}%
  \BibitemOpen
  \bibfield  {author} {\bibinfo {author} {\bibfnamefont {N.}~\bibnamefont
  {Read}},\ }\href {\doibase 10.1103/PhysRevB.79.045308} {\bibfield  {journal}
  {\bibinfo  {journal} {Phys. Rev. B}\ }\textbf {\bibinfo {volume} {79}},\
  \bibinfo {pages} {45308} (\bibinfo {year} {2009})}\BibitemShut {NoStop}%
\bibitem [{\citenamefont {Read}\ and\ \citenamefont {Rezayi}(2011)}]{Read2011}%
  \BibitemOpen
  \bibfield  {author} {\bibinfo {author} {\bibfnamefont {N.}~\bibnamefont
  {Read}}\ and\ \bibinfo {author} {\bibfnamefont {E.~H.}\ \bibnamefont
  {Rezayi}},\ }\href {\doibase 10.1103/PhysRevB.84.085316} {\bibfield
  {journal} {\bibinfo  {journal} {Phys. Rev. B}\ }\textbf {\bibinfo {volume}
  {84}},\ \bibinfo {pages} {85316} (\bibinfo {year} {2011})}\BibitemShut
  {NoStop}%
\bibitem [{\citenamefont {Hoyos}\ \emph {et~al.}(2013)\citenamefont {Hoyos},
  \citenamefont {Moroz},\ and\ \citenamefont {Son}}]{Hoyos2013}%
  \BibitemOpen
  \bibfield  {author} {\bibinfo {author} {\bibfnamefont {C.}~\bibnamefont
  {Hoyos}}, \bibinfo {author} {\bibfnamefont {S.}~\bibnamefont {Moroz}}, \ and\
  \bibinfo {author} {\bibfnamefont {D.~T.}\ \bibnamefont {Son}},\ }\href
  {\doibase 10.1103/PhysRevB.89.174507} {\bibfield  {journal} {\bibinfo
  {journal} {Phys. Rev. B}\ }\textbf {\bibinfo {volume} {89}},\ \bibinfo
  {pages} {174507} (\bibinfo {year} {2013})}\BibitemShut {NoStop}%
\bibitem [{\citenamefont {Wen}\ and\ \citenamefont
  {Zee}(1992{\natexlab{b}})}]{Wen1992}%
  \BibitemOpen
  \bibfield  {author} {\bibinfo {author} {\bibfnamefont {X.~G.}\ \bibnamefont
  {Wen}}\ and\ \bibinfo {author} {\bibfnamefont {A.}~\bibnamefont {Zee}},\
  }\href {http://journals.aps.org/prl/abstract/10.1103/PhysRevLett.69.953}
  {\bibfield  {journal} {\bibinfo  {journal} {Phys. Rev. Lett.}\ }\textbf
  {\bibinfo {volume} {69}},\ \bibinfo {pages} {953} (\bibinfo {year}
  {1992}{\natexlab{b}})}\BibitemShut {NoStop}%
\bibitem [{\citenamefont {Cooper}(2008)}]{Cooper2008}%
  \BibitemOpen
  \bibfield  {author} {\bibinfo {author} {\bibfnamefont {N.~R.}\ \bibnamefont
  {Cooper}},\ }\href@noop {} {\bibfield  {journal} {\bibinfo  {journal} {Adv.
  Phys.}\ }\textbf {\bibinfo {volume} {57}},\ \bibinfo {pages} {539} (\bibinfo
  {year} {2008})}\BibitemShut {NoStop}%
\bibitem [{\citenamefont {Chan}\ \emph {et~al.}(2013)\citenamefont {Chan},
  \citenamefont {Hughes}, \citenamefont {Ryu},\ and\ \citenamefont
  {Fradkin}}]{Chan2013}%
  \BibitemOpen
  \bibfield  {author} {\bibinfo {author} {\bibfnamefont {A.}~\bibnamefont
  {Chan}}, \bibinfo {author} {\bibfnamefont {T.~L.}\ \bibnamefont {Hughes}},
  \bibinfo {author} {\bibfnamefont {S.}~\bibnamefont {Ryu}}, \ and\ \bibinfo
  {author} {\bibfnamefont {E.}~\bibnamefont {Fradkin}},\ }\href {\doibase
  10.1103/PhysRevB.87.085132} {\bibfield  {journal} {\bibinfo  {journal} {Phys.
  Rev. B}\ }\textbf {\bibinfo {volume} {87}},\ \bibinfo {pages} {085132}
  (\bibinfo {year} {2013})}\BibitemShut {NoStop}%
\bibitem [{\citenamefont {Teo}\ \emph {et~al.}(2015)\citenamefont {Teo},
  \citenamefont {Hughes},\ and\ \citenamefont {Fradkin}}]{Teo2015}%
  \BibitemOpen
  \bibfield  {author} {\bibinfo {author} {\bibfnamefont {J.~C.~Y.}\
  \bibnamefont {Teo}}, \bibinfo {author} {\bibfnamefont {T.~L.}\ \bibnamefont
  {Hughes}}, \ and\ \bibinfo {author} {\bibfnamefont {E.}~\bibnamefont
  {Fradkin}},\ }\href {\doibase 10.1016/j.aop.2015.05.012} {\bibfield
  {journal} {\bibinfo  {journal} {Ann. Phys.}\ }\textbf {\bibinfo {volume}
  {360}},\ \bibinfo {pages} {349} (\bibinfo {year} {2015})}\BibitemShut
  {NoStop}%
\end{thebibliography}%

\end{document}